\documentclass[pra,twocolumn,aps,superscriptaddress,showpacs,floatfix]{revtex4}
\usepackage[latin9]{inputenc}
\usepackage{amsmath}
\usepackage{amssymb}
\usepackage{graphicx}
\usepackage{multirow}
\usepackage{float}
\usepackage{tabularx}
\usepackage{makecell}
\usepackage{bm}
\makeatletter

\@ifundefined{textcolor}{}
{%
 \definecolor{BLACK}{gray}{0}
 \definecolor{WHITE}{gray}{1}
 \definecolor{RED}{rgb}{1,0,0}
 \definecolor{GREEN}{rgb}{0,1,0}
 \definecolor{BLUE}{rgb}{0,0,1}
 \definecolor{CYAN}{cmyk}{1,0,0,0}
 \definecolor{MAGENTA}{cmyk}{0,1,0,0}
 \definecolor{YELLOW}{cmyk}{0,0,1,0}
}
\newcommand{\tabincell}[2]{\begin{tabular}{@{}#1@{}}#2\end{tabular}}

\usepackage{amsthm}

\newtheorem*{prop*}{Proposition}
\makeatother

\begin{document}
\title{Fisher information under decoherence in Bloch representation}

\author{Wei Zhong}
\affiliation{Advanced Science Institute, RIKEN, Wako-shi, Saitama 351-0198, Japan}
\affiliation{Zhejiang Institute of Modern Physics, Department of Physics, Zhejiang University, Hangzhou 310027, China}

\author{Zhe Sun}
\affiliation{Advanced Science Institute, RIKEN, Wako-shi, Saitama 351-0198, Japan}
\affiliation{Department of Physics, Hangzhou Normal University, Hangzhou 310036, China}

\author{Jian Ma}
\affiliation{Advanced Science Institute, RIKEN, Wako-shi, Saitama 351-0198, Japan}
\affiliation{Zhejiang Institute of Modern Physics, Department of Physics, Zhejiang University, Hangzhou 310027, China}

\author{Xiaoguang Wang}
\email{xgwang@zimp.zju.edu.cn}
\affiliation{Advanced Science Institute, RIKEN, Wako-shi, Saitama 351-0198, Japan}
\affiliation{Zhejiang Institute of Modern Physics, Department of Physics, Zhejiang University, Hangzhou 310027, China}

\author{Franco Nori}
\affiliation{Advanced Science Institute, RIKEN, Wako-shi, Saitama 351-0198, Japan}
\affiliation{Physics Department, The University of Michigan, Ann Arbor, MI 48109-1040, USA}

\begin{abstract}
The dynamics of two variants of quantum Fisher information under decoherence are investigated from a geometrical point of view.
We first derive the explicit formulas of these two quantities for a single qubit in terms of the Bloch vector.
Moreover, we obtain analytical results for them under three different decoherence channels, which are expressed as affine transformation matrices.
Using the hierarchy equation method, we numerically study the dynamics of both the two information in a dissipative model
and compare the numerical results with the analytical ones obtained by applying the rotating-wave approximation.
We further express the two information quantities in terms of the Bloch vector for a qudit,
by expanding the density matrix and Hermitian operators in a common set of generators of the Lie algebra $\mathfrak{su}(d)$.
By calculating the dynamical quantum Fisher information, we find that the collisional dephasing significantly
diminishes the precision of phase parameter with the Ramsey interferometry.
\end{abstract}

\maketitle

\section{Introduction}

Quantum Fisher information (QFI), which is one of the most important
quantities for both quantum estimation theory and quantum information
theory, has been widely studied \cite{Petz1996,Smerzi2009,Sun2010,Ma2009,Luis2010,Ma2011,Ma2011Rep,Toth2012,Smerzi2012}.
In the field of quantum estimation, the main task is to determine the
value of an unknown parameter labeling the quantum system, and a primary goal is to enhance the precision of the resolution
\cite{Giovannetti2006,Giovannetti2011,Huegla1997,Escher2011,Uys2007,Shaji2007,Boixo2008,Roy2008,Pairs2009,You2011,Joo2011}.
The inverse of the QFI provides the lower bound of error of the estimation \cite{HelstromBook,HolevoBook}.
Hence, how to increase the QFI become the key problem to be solved.
Moreover, the QFI can be used to measure the statistical distinguishability on the space of the density operators in the quantum information geometry \cite{Wootters1981,Braunstein1994}.
Recently, the QFI flow was proposed as a quantitative measure of the
information flow and provides a novel perspective on observing the
non-Markovian behavior in open quantum systems \cite{Lu2010}.

There are several variants of quantum versions of Fisher information,
among which the one based on the symmetric logarithmic derivative (SLD) operator has been used most widely and possesses many good properties \cite{HelstromBook,HolevoBook}, such as convexity, remaining invariant under the unitary evolution, 
and the total amount of the QFI equivalent to the summation of the QFIs of all subsystem being uncorrelated. 
In this paper, we also study another variant of the quantum version of
the Fisher information, which is closely related with the skew information \cite{WignerYanase1963}.
The skew information was proposed to measure the amount of information
that a quantum state contains with respect to the observable which does not commute with additive conserved quantities, such as Hamiltonians,
momenta \cite{WignerYanase1963}.
It is also used to measure the quantum uncertainty \cite{Luo2006} and
quantify quantum correlations in bipartite state in a recent work \cite{Luo2012}.
In the following, we shall show that the two QFIs formally possess many similar features \cite{Petz1996}, but they are different not only for the concrete expressions of the definitions, but also for their applications.
As is mentioned previously, the QFI based on SLD operator is mainly applied in the quantum metrology, however, the variant version of the QFI plays an important role in quantum state discrimination \cite{Calsamiglia2008,Audenaert2007}.

From the information theory, both two different QFIs characterize
the content of information contained in the quantum systems.
A crucial property of these two information quantities is that
they decrease monotonically under the completely positive and
trace-preserving (CPT) maps~\cite{Stinespring1955,Kraus1971,Gibilisco2003,Petz1996}.
The monotonicity property manifests the information loss under the CPT map.
Just as the introduction of the QFI flow \cite{Lu2010}, can we observe
the non-Markovian properties from the other information quantities perspective?

Here, we will address this problem by calculating these two QFIs for a
single qubit and derive the explicit formulas in the Bloch representation,
which greatly facilitates the computing of these two quantities.
The main results in this paper are that the dynamical QFIs, in
the presence of decoherences, are analytically solved.
Here, the irreversible processes are modeled via three decoherence
channels \cite{Ma2011Rep,NielsenBook,Wang2010,Bartkowiak2011}:
phase-damping channel~(PDC), depolarizing channel~(PDC), and the
generalized amplitude damping channel~(GADC) \cite{Aolita2008}.
The analytical results for the two information quantities under
those channels are obtained.
We will show that the values of the two QFIs monotonically decrease
with time, apart from the isolated case that the QFI based on the
SLD operator about the amplitude parameter $\theta$ remain invariant
under the PDC.
In order to further identify the behaviors of the two QFIs subject
to quantum noise, we discuss a simple model of a two-level system
coupled to a reservoir with a Lorentzian spectral density \cite{BreuerBook,Gorini1976}.
By using the hierarchy equation method \cite{Shao,Yan,Tanimura,Tanimura2006,Ishizaki,Ishizaki&Fleming,Ma2012,Yin2012},
we numerically analyze the dynamics of the two QFIs, and compare
these with the analytical solutions by using the rotating-wave approximation (RWA).
We further generalize the results to the qudit system. Meanwhile,
we verify that they are also applicable for the $N$-qubit system
with symmetry exchange. For the sake of clarity, we calculate the
dynamical quantum Fisher information in the presence of collisional dephasing.

This paper is organized as follows.
In Sec.~\ref{QFS}, we first review two different definitions of the QFI,
and give the explicit formulae for the QFIs for a single qubit system.
In Sec.~\ref{QFI_qubit}, we obtain the analytical results for the two
quantities under three different decoherence channels, and the numerical
results are given. Moreover, in Sec.~\ref{QFI_qudit}(a), we generalize
the expressions of the two QFIs for a qudit system, and the QFI in a
noisy environment for an $N$-qubit system is discussed in Sec.~\ref{QFI_qudit}(b).
Finally, the conclusion are given in Sec.~\ref{conclusion}.

\section{Fisher information}\label{QFS}

In this section, we briefly summarize two variants of definition of the
QFIs \cite{WignerYanase1963,Fisher1925},
which are referred to as two different extensions from the classical
Fisher information. We also discuss the relations between the QFI and
the Bures distance \cite{Braunstein1994,Bures1969,Uhlmann1976,Hubner1992}
as well as the QFI and the Hellinger distance \cite{Luo2004}.
In the Bloch representation, we derive the explicit formulas of these
two information quantities for the single qubit system.

\subsection{Fisher information and Bures distance}

The classical Fisher information, originating
from the statistical inference, is a way of
measuring the amount of information that an
observable random variable $X$ carries about
an unknown parameter $\lambda$.
Suppose that $\left\{p_{i}\left(\lambda\right),\,\lambda\in\mathbb{R}\right\}_{i=1}^{N}$
is the probability density conditioned on the
fixed value of the parameter $\lambda=\lambda^{*}$
with measurement outcomes $\left\{x_{i}\right\}$.
The classical Fisher information is defined as
\begin{equation}
  F_{\lambda} = \sum_{i}p_{i}\left(\lambda\right)\left[\frac{\partial\ln p_{i}\left(\lambda\right)}{\partial\lambda}\right]^{2},
  \label{eq:FI1}
\end{equation}
which characterizes the inverse variance of the
asymptotic normality of a maximum-likelihood estimator.
Here we have assumed that the observable $\hat{X}$
is a discrete variable. If it is continuous, the
summation in Eq.~\eqref{eq:FI1} should be replaced
by an integral.

The quantum analog of the Fisher information is
formally generalized from Eq.~\eqref{eq:FI1} and
defined as
\begin{equation}
  \mathcal{F}_{\lambda}={\rm Tr}\left(\rho_\lambda\,L_\lambda^{2}\right)
  ={\rm Tr}\left[\left(\partial_{\lambda}\,\rho_{\lambda}\right)\,L_\lambda\right],
  \label{eq:QFI1}
\end{equation}
in terms of the symmetric logarithmic derivative
(SLD) operator $L_\lambda$, which is a Hermitian
operator determined by
\begin{equation}
  \partial_{\lambda}\,\rho_{\lambda}=\frac{1}{2}\left\{\rho_\lambda,L_\lambda\right\},
  \label{eq:SLD}
\end{equation}
where $\partial_{\lambda}\equiv\frac{\partial}{\partial\lambda}$
and $\left\{\,\cdot\,,\,\cdot\,\right\}$
denotes the anticommutator.
By diagonalizing the density matrix as
$\rho_\lambda=\sum_{i}\varrho_{i}\left|\psi_{i}\right\rangle\left\langle \psi_{i}\right|$,
associated with $\varrho_{i}\geq0$ and
$\sum_{i}\varrho_{i}=1$, the elements
of the SLD operator are completely defined
under the condition $\varrho_{i}+\varrho_{j}\neq0$.
Therefore, Eq.~\eqref{eq:QFI1} can be
expressed as
\begin{equation}
  \mathcal{F}_{\lambda}=
  \sum_{i'}\frac{\left(\partial_{\lambda}\varrho_{i'}\right)^{2}}{\varrho_{i'}}
  +2\sum_{i\neq j}\frac{\left(\varrho_{i}-\varrho_{j}\right)^{2}}{\varrho_{i}+\varrho_{j}}
  \left|\left\langle\psi_{i}|\partial_{\lambda}\psi_{j}\right\rangle \right|^{2},
  \label{eq:QFI1_sum}
\end{equation}
where the first and the second summations
involve sums over all $\varrho_{i'}\neq0$ and
$\varrho_{i}+\varrho_{j}\neq0$, respectively.
In Eq.~\eqref{eq:QFI1_sum}, the first term is
equal to the classical Fisher information of
Eq.~\eqref{eq:QFI1}, which is called the
classical term, and the second term is called
the quantum term. For pure states, Eq.~\eqref{eq:QFI1_sum}
reduces to
\begin{equation}
  \mathcal{F}_{\lambda}=
  4\left[\left\langle \partial_{\lambda}\psi|\partial_{\lambda}\psi\right\rangle
  -\left|\left\langle\psi|\partial_{\lambda}\psi\right\rangle \right|^{2}\right].
  \label{eq:QFI1_sum_pure}
\end{equation}
An essential feature of the QFI is that we can
obtain the achievable lower bound of the mean-square
error of the unbiased estimator for the parameter
$\lambda$, i.e., the so-called quantum Cram\'er-Rao
(QCR) theorem:
\begin{equation}
  {\rm Var}\big{(}\hat{\lambda}\big{)}\geq\frac{1}{\nu\mathcal{F}_{\lambda}},
  \label{eq:QCRbound}
\end{equation}
where ${\rm Var}\left(\cdot\right)$ denotes the
variance, $\hat{\lambda}$ denotes the unbiased
estimator, and $\nu$ represents the number of
repeated experiments.

As shown in a seminal work, see Ref.~\cite{Braunstein1994},
they observed that the estimability of a set of
parameters parameterizing the family of the quantum
states $\left\{\rho_{\lambda}\right\}$, which is
characterized by the QFI, is naturally related to
the distinguishability of the states on the manifold
of the quantum states, which is measured by the
Bures distance. They also proved that the QFI is
simply proportional to the Bures distance \cite{Braunstein1994}
\begin{equation}
  D_{{\rm B}}^{2}\left[\rho\left(\lambda\right),\rho\left(\lambda+d\lambda\right)\right]
  =\frac{1}{4}\mathcal{F}_{\lambda}\,d\lambda^{2}.
  \label{eq:BuresDistance_QFI}
\end{equation}
The Bures distance measures the distance between
two quantum states, and is defined as \cite{Bures1969,Uhlmann1976,Hubner1992}
\begin{equation}
  D_{{\rm B}}^{2}\left(\rho,\sigma\right):=2\left(1-{\rm Tr}\sqrt{\rho^{1/2}\sigma\rho^{1/2}}\right),
  \label{eq:BuresDistance}
\end{equation}
where the second term in the bracket is the
so-called Uhlmann fidelity \cite{Uhlmann1976}.
Meanwhile, the explicit formula of the QFI for
the two-dimensional density matrices is obtained
by \cite{Dittmann1999}
\begin{equation}
  \mathcal{F}_{\lambda}={\rm Tr}\left(\partial_{\lambda}\rho\right)^{2}+
  \frac{1}{{\rm det}\rho}{\rm Tr}\left(\rho\,\partial_{\lambda}\rho\right)^{2}.
  \label{eq:QFI_Dittmann}
\end{equation}

In the Bloch sphere representation, any qubit
state can be written as
\begin{equation}
  \rho=\frac{1}{2}\left(\openone+\bm{\omega}\cdot\hat{\bm{\sigma}}\right),\label{eq:qubit_BlochRep}
\end{equation}
where $\bm{\omega}=\left(\omega_{x},\,\omega_{y},\,\omega_{z}\right)^{{\rm T}}$
is the real Bloch vector and $\hat{\bm{\sigma}}=\left(\hat{\sigma}_{x},\,\hat{\sigma}_{y},\,\hat{\sigma}_{z}\right)$
denotes the Pauli matrices.
Apparently, the eigenvalues of the density
operator are $\left(1\pm \omega\right)/2$,
with the length of the Bloch vector $\omega\equiv\left|\bm{\omega}\right|$.
Here, the Blcoh vector satisfies $\omega\leq1$,
the equality holds for pure states.
In the Bloch representation, $\mathcal{F}_{\lambda}$
can be represented as follows
\begin{equation}
  \mathcal{F}_{\lambda}=
  \begin{cases}
  \left|\partial_{\lambda}\bm{\omega}\right|^{2}+
  \frac{\left(\bm{\omega}\cdot\partial_{\lambda}\bm{\omega}\right)^{2}}
  {1-\left|\bm{\omega}\right|^{2}}, & \omega<1,\\
  \left|\partial_{\lambda}\bm{\omega}\right|^{2}, & \omega=1.\label{eq:QFI1_qubit}
\end{cases}
\end{equation}
The first line of the above equation is only
applicable for the mixed states, which can be
straightly obtained by substituting Eq.~\eqref{eq:qubit_BlochRep}
into Eq.~\eqref{eq:QFI_Dittmann}.

For pure states, we have equation $\rho^{2}=\rho$.
Taking differential on both side of this equation
with respect to $\lambda$, one gets
\begin{equation}
  \partial_{\lambda}\rho=\partial_{\lambda}\rho^{2}
  =\rho\left(\partial_{\lambda}\rho\right)+\left(\partial_{\lambda}\rho\right)\rho.
  \label{eq:partial_rho}
\end{equation}
The SLD operator is given as
\begin{equation}
  L=2\,\partial_{\lambda}\rho, \label{eq:SLD_pure}
\end{equation}
by comparing Eq.~\eqref{eq:partial_rho} with Eq.~\eqref{eq:SLD}.
Substituting Eqs.~\eqref{eq:qubit_BlochRep} and
\eqref{eq:SLD_pure} into Eq.~\eqref{eq:QFI1},
and using the relation
\begin{equation}
  {\rm Tr}\left[\left(\bm{a}\cdot\hat{\bm{\sigma}}\right)\left(\bm{b}\cdot\hat{\bm{\sigma}}\right)\right]
  =2\,\bm{a}\cdot\bm{b}
  \label{eq:Tr_relation}
\end{equation}
finally yield the second line of Eq.~\eqref{eq:QFI1_qubit},
i.e., $\mathcal{F}_{\lambda}$ for pure states is
the norm of the derivative of the Bloch vector.

\subsection{Fisher information and Hellinger distance}

By extending Eq.~\eqref{eq:FI1} into the quantum
regime in a different way, we will obtain a variant QFI \cite{Luo2003}.
Eq.~\eqref{eq:FI1} can be equivalently expressed as
\begin{equation}
  F_{\lambda}=4\sum_{i}\left(\partial_{\lambda}\sqrt{p_{i}\left(\lambda\right)}\right)^{2}.
  \label{eq:FI2}
\end{equation}
By straightly replacing the summation by a trace,
the probability $p_{i}\left(\lambda\right)$ by a
density matrix $\rho_{\lambda}$, and the differential
$\partial_{\lambda}$ by the inner differential $\partial_{\lambda}\cdot\equiv i\left[\hat{G},\cdot\right]$
with $\hat{G}$ being a fixed self-adjoint operator
and $[\,\cdot,\,\cdot\,]$ denoting the commutator.
One obtains the following equation
\begin{equation}
  \mathcal{I}_\lambda:=4{\rm Tr}\left(\partial_\lambda\sqrt\rho_\lambda\right)^{2}=-4{\rm Tr}\left[\rho^{1/2},\hat{G}\right]^{2}.
  \label{eq:QFI2}
\end{equation}
In the particular case, suppose that $\rho_{\lambda}\equiv e^{-i\hat{G}\lambda}\rho e^{i\hat{G}\lambda}$,
i.e., $\rho_\lambda$ satisfies the Landau-von Neumann
equation $i\partial_{\lambda}\rho_{\lambda}=\left[\hat{G},\rho_{\lambda}\right]$.
Note that here we define Eq.~\eqref{eq:QFI2} as a paradigmatic version of the QFI.
Actually, Eq.~\eqref{eq:QFI2} is the so-called skew information
\begin{equation}
  \mathcal{I}_{\rm WY}:=-\frac{1}{2}{\rm Tr}\left[\rho_\lambda^{1/2},\hat{G}\right]^{2},
  \label{eq:SI}
\end{equation}
which is introduced by Wigner and Yanase \cite{WignerYanase1963},
with ignorance of a negligible constant number here,
i.e. $\mathcal{I}=8\mathcal{I}_{\rm WY}$.
The skew information (SI) is a measure of the information
contained in a quantum state $\rho_{\lambda}$
with respect of a fixed conserved observable $\hat{G}$.

By inserting the spectrum decomposition
$\rho=\sum_{i}\varrho_{i}\left|\psi_{i}\right\rangle\left\langle \psi_{i}\right|$
into Eq.~\eqref{eq:QFI2}, one obtain
\begin{equation}
  \mathcal{I}_{\lambda}=\sum_{i'}\frac{\left(\partial_{\lambda}\varrho_{i'}\right)^{2}}{\varrho_{i'}}+4\sum_{i\neq j}\left(\sqrt{\varrho_{i}}-\sqrt{\varrho_{j}}\right)^{2}\left|\left\langle \psi_{i}|\partial_{\lambda}\psi_{j}\right\rangle \right|^{2},
  \label{eq:QFI2_sum}
\end{equation}
where the first summation is over all $\varrho_{i}\neq0$,
the same requirement as in Eq.~\eqref{eq:QFI2_sum}.
Comparing with Eq.~\eqref{eq:QFI1_sum}, the classical terms
in Eqs.~\eqref{eq:QFI1_sum} and \eqref{eq:QFI2_sum} are the same,
but the quantum terms are different.
For pure states, Eq.\,\eqref{eq:QFI2_sum} reduces to
\begin{equation}
  \mathcal{I}_{\lambda}=8\left[\left\langle \partial_{\lambda}\psi|\partial_{\lambda}\psi\right\rangle -\left|\left\langle\psi|\partial_{\lambda}\psi\right\rangle \right|^{2}\right],\label{eq:QFI2_sum_pure}
\end{equation}
which is twice as much as the QFI of Eq.~\eqref{eq:QFI1}
given in Eq.~\eqref{eq:QFI1_sum_pure}.
The factor-of-$2$ difference between $\mathcal{F}_\lambda$ and $\mathcal{I}_\lambda$
result from their different coefficients (or weights) of the
quantum terms in Eqs.~\eqref{eq:QFI1_sum} and \eqref{eq:QFI2_sum}.
For pure states, the classical terms vanish, and only the quantum terms contribute.
Then we obtain Eqs.~\eqref{eq:QFI1_sum_pure} and \eqref{eq:QFI2_sum_pure}.

Similar to the relation between the QFI of Eq.~\eqref{eq:QFI1}
and the Bures distance given in Eq.~\eqref{eq:BuresDistance_QFI},
there formally exists the same relation between the QFI of Eq.~\eqref{eq:SI}
and quantum Hellinger distance \cite{Luo2004}:
\begin{equation}
  D_{{\rm QH}}^{2}\left[\rho\left(\lambda\right),\rho\left(\lambda+d\lambda\right)\right]
  =\frac{1}{4}\mathcal{I}_{\lambda}\,d\lambda^{2}.\label{eq:HellingerDistance_SI}
\end{equation}
The latter is the quantum version of the classical Hellinger distance,
which measures the distance between two probability distributions.
In the space of the quantum state, the quantum Hellinger distance is defined as
\begin{equation}
  D_{{\rm QH}}^{2}\left(\rho,\sigma\right):=2\left(1-{\rm Tr}\sqrt{\rho}\sqrt{\sigma}\right),
\end{equation}
where the last term in the bracket is called quantum affinity \cite{Luo2004}.

For a $2\times2$ density matrix, we derive the explicit formula
of the QFI $ \mathcal{I}_{\lambda}$ as follows:
\begin{eqnarray}
  \mathcal{I}_{\lambda} &=&
  \alpha\,{\rm Tr}\left(\partial_{\lambda}\rho\right)^{2}-\beta\,{\rm Tr}\left(\rho\,\partial_{\lambda}\rho\right)^{2},
  \label{eq:SI_Dittmann}
\end{eqnarray}
where the coefficients are determined by
\begin{eqnarray}
  \alpha & = & \frac{1}{1-4\det\rho}\left[\frac{4\left(1-2\det\rho\right)}{\left(1+2\sqrt{\det\rho}\right)}-1\right],\label{eq:coef_alpha}\\
  \beta & = & \frac{1}{1-4\det\rho}\left(\frac{8}{1+2\sqrt{\det\rho}}-\frac{1}{\det\rho}\right)\label{eq:coef_beta}.
\end{eqnarray}

In the Bloch representation, $\mathcal{I}_{\lambda}$
can be represented as follows
\begin{equation}
  \mathcal{I}_{\lambda}=
  \begin{cases}
  \frac{2\left|\partial_{\lambda}\bm{\omega}\right|^{2}}{1+\sqrt{1-\left|\bm{\omega}\right|^{2}}}
  +\Theta_{\bm{\omega}}\left(\bm{\omega}\cdot\partial_{\lambda}\bm{\omega}\right)^{2}, & \omega<1,\\
  2\left|\partial_{\lambda}\bm{\omega}\right|^{2}, & \omega=1,\label{eq:QFI2_qubit}
\end{cases}
\end{equation}
where the coefficient is given by
\begin{equation}
  \Theta_{\bm{\omega}}=\frac{1}{1-\left|\bm{\omega}\right|^{2}}-\frac{1}{\left(1+\sqrt{1-\left|\bm{\omega}\right|^{2}}\right)^{2}}.
  \label{eq:gamma_coefs}
\end{equation}
The first line of Eq.~\eqref{eq:QFI2_qubit} is only applicable for mixed states,
which can be directly derived by substituting Eq.~\eqref{eq:qubit_BlochRep} into
Eqs.~\eqref{eq:SI_Dittmann},~\eqref{eq:coef_alpha} and \eqref{eq:coef_beta}.

For pure states, we have
\begin{equation}
  \sqrt{\rho}=\rho.\label{eq:sqrt_rho1}
\end{equation}
With Eqs.~\eqref{eq:Tr_relation} and \eqref{eq:sqrt_rho1},
one can obtain the second line of Eq.~\eqref{eq:QFI2_qubit} from Eq.~\eqref{eq:QFI2},
i.e., $\mathcal{I}_{\lambda}$ for pure states is the norm of
the derivative of the Bloch vector up to a factor of $2$.

\subsection{Example}\label{ini_state}

Having obtained Eqs.~\eqref{eq:QFI1_qubit} and \eqref{eq:QFI2_qubit},
we consider a simple example to calculate the two variant QFIs
for a pure state with different parameters.
We adopt the standard notation where
$\left|1\right\rangle\equiv \left|\downarrow\right\rangle$ and
$\left|0\right\rangle\equiv \left|\uparrow\right\rangle$
correspond to the ground state and excited state, respectively.
Consider an arbitrary single-qubit state
\begin{equation}
  \left|\psi\right\rangle =\cos\frac{\theta}{2}\left|0\right\rangle +e^{i\phi}\sin\frac{\theta}{2}\left|1\right\rangle,
  \label{eq:ini_state}
\end{equation}
of which the Bloch vector is denoted as
$\bm{\omega}=\left(\sin\theta\cos\phi,\,\sin\theta\sin\phi,\,\cos\theta\right)^{\rm T}$,
with $\theta$ and $\phi$ referring to the polar and azimuth angles on the Bloch sphere.
Here, the two parameters $\theta$ and $\phi$ in Eq.~\eqref{eq:ini_state}
are assumed to be unitary encoded.

Now we compute the QFIs of the single qubit state
Eq.~\eqref{eq:ini_state} in terms of the parameters:
$\theta$ (amplitude parameter) and $\phi$ (phase parameter).
With the help of Eq.\ \eqref{eq:QFI1_qubit},
one can easily obtain $\mathcal{F}_{\theta}=1$ and $\mathcal{F}_{\phi}=\sin^{2}\theta$.
For the amplitude parameter $\theta$,
$\mathcal{F}_{\theta}$ is constantly equal to $1$,
independent of both two parameters $\theta$ and $\phi$.
While $\mathcal{F}_{\phi}$ about $\phi$ depends on
the parameter $\theta$ and reaches the maximum value $1$,
when $\theta=\pi/2$, i.e., $\left|\psi\right\rangle $ is a equal-weighted state
$\left|\psi\right\rangle=\left(\left|0\right\rangle +e^{i\phi}\left|1\right\rangle\right)/\sqrt{2}$ \cite{Giovannetti2006}.
According to Eq.~\eqref{eq:QFI2_qubit}, we obtain $\mathcal{I}_{\theta}=2$ and $\mathcal{I}_{\phi}=2\sin^{2}\theta$.
Similar to $\mathcal{F}_{\theta}$, $\mathcal{I}_{\theta}$ is also
independent of the parameters $\theta$ and $\phi$.
Meanwhile, $\mathcal{I}_{\phi}$ reaches the maximum
value $2$ at the point $\theta=\pi/2$.
In Sec.~\ref{num}, we will assume that the qubit is initially
in the equally weighted superposition of the two states
$\left|0\right\rangle$ and $\left|1\right\rangle$,
so both $\mathcal{F}_{\phi}$ and $\mathcal{I}_{\phi}$
reach their maximum values.

\section{QFI for a single qubit under Decoherence}\label{QFI_qubit}

Decoherence occurs when a quantum system interacts with its environment,
and it is unavoidable in almost all the realistic quantum systems.
A quantum noisy dynamical process can be generally described
by a map $\mathcal{E}$, using the Kraus representation
\begin{equation}
  \mathcal{E}(\rho)=\sum_{\mu}K_{\mu}\,\rho\,K_{\mu}^{\dagger},\label{eq:Kraus_map}
\end{equation}
where $K_{\mu}$ are the Kraus operators satisfying
$\sum_{\mu}K_{\mu}^{\dagger}\,K_{\mu}=\openone$,
which leads to the map $\mathcal{E}$ being a CPT map
\cite{Stinespring1955,Kraus1971}.
In the Bloch representation, such an non-unital
trace-preserving process can be represented by
an affine map on the generalized Bloch vector \cite{Ueda2010,NielsenBook}.

For qubit system, Eq.~\eqref{eq:Kraus_map} can be equivalently represented as
\begin{equation}
  \mathcal{E}(\rho)=\frac{1}{2}\openone+\frac{1}{2}\left(A\,\bm{\omega}+\bm{{\rm c}}\right)\cdot\hat{\bm{\sigma}},
  \label{eq:affine_map_qubit}
\end{equation}
where $A$ is a $3\times3$ real transformation matrix with
elements defined as $A_{ij}=\frac{1}{2}{\rm Tr}\left[\sigma_i\mathcal{E}\left(\sigma_j\right)\right]$,
and $\bm{c}\in\mathbb{R}^{3}$ is the translation vector with
elements given by $c_{i}=\frac{1}{2}{\rm Tr}\left[\sigma_i\mathcal{E}\left(\openone\right)\right]$.
From Eqs.~\eqref{eq:qubit_BlochRep} and \eqref{eq:affine_map_qubit},
it indicates that under the decoherence, the Bloch vector $\bm{\omega}$
of Eq.~\eqref{eq:qubit_BlochRep} is mapped as $\mathcal{E}(\,\bm{\omega}):=A\,\bm{\omega}+\bm{c}$.
In parameter estimation, the unknown parameter is generally
encoded on the probes through a unitary or non-unitary evolution \cite{Escher2011}.
In this paper, we do not consider the case of the non-unitary
parametrization. $A$ and $\bm{c}$ are assumed to be parameter-independent,
i.e., the decoherence process will not introduce the parameter.

With Eq.~\eqref{eq:QFI1_qubit}, the dynamic of the QFI based on
the SLD operator for a single qubit under decoherence channels
can be generally expressed as
\begin{equation}
  \mathcal{F}_\lambda = \left|\partial_{\lambda}\mathcal{E}(\,\bm{\omega})\right|^2+
  \frac{\left[\mathcal{E}(\,\bm{\omega})\cdot\partial_\lambda\mathcal{E}(\,\bm{\omega})\right]^2}{1-\left|\mathcal{E}(\,\bm{\omega})\right|^2}.
  \label{eq:QFI1_noisy}
\end{equation}
Similarly, according to Eq.~\eqref{eq:QFI2_qubit},
the dynamic of the variant version of the QFI can be written as
\begin{equation}
   \mathcal{I}_{\lambda} =
   \frac{2\left|\partial_{\lambda}\mathcal{E}(\,\bm{\omega})\right|^{2}}{1+\sqrt{1-\left|\mathcal{E}(\,\bm{\omega})\right|^{2}}}
   +\Theta_{\mathcal{E}(\bm{\omega})}\left[\mathcal{E}(\,\bm{\omega})\cdot\partial_{\lambda}\mathcal{E}(\,\bm{\omega})\right]^{2},
   \label{eq:QFI2_noisy}
\end{equation}
Given an input state $\left|\psi\right\rangle$,
the dynamics of the two QFIs under quantum channels are fully
determined by the affine transformation matrix $A$ and the
translation vector $\bm{c}$. It is noted that Eqs.~\eqref{eq:QFI1_noisy}
and \eqref{eq:QFI2_noisy} are the general results that are
applicable to all of those cases with different parametrization
processes.

\subsection{Dynamics of the QFIs under three decoherence channels}\label{single_qubit}

Below, we will study the dynamics of the two variant versions of
the QFI under three paradigmatic types of quantum channels
\cite{NielsenBook,Aolita2008}: phase-damping channel (PDC),
depolarizing channel (DPC), and generalized amplitude-damping channel (GADC)
modeling a thermal bath at arbitrary temperature, which will be
reduced to the purely dissipative amplitude-damping channel (ADC)
when the environment temperature becomes zero.
These channels are the prototype models of dissipation relevant
in various experimental systems,

\begin{table*}[tbp!]
  \caption{Analytical results for the time-evolutions of the two QFI
  quantities in terms of the parameters $\theta$ and $\phi$ for the
  single qubit state $\left|\psi\right\rangle$ in Eq.~\eqref{eq:ini_state}
   under the quantum decoherence channels.
  The decoherence channels can geometrically be described as affine maps,
  and the transformation matrices $A$ and the translation vectors
  $\bm{c}$ for each channel are listed.
  In order to simplify the expression of the results,
  we set $X=\sin\theta\sqrt{1-s^{2}}$,
  $Y=2\cos^{2}\frac{\theta}{2}\sqrt{s\left(1-s\right)}$,
  $Z=\sqrt{1-\bar{s}\sin^{2}\theta-\left[\bar{p}\left(\alpha-\beta\right)-\bar{s}\cos\theta\right]^{2}}$,
  $R_{1}=\bar{s}\left[1+\bar{s}+\bar{p}\cos\left(2\,\theta\right)\right]$, and
  $R_{2}=\bar{p}^{2}\bar{s}^{2}\left(\alpha-\beta+\cos\theta\right)^{2}\sin^{2}\theta$.}
  \label{tab:Table1}
  \centering{}
  \begin{ruledtabular}
  \begin{tabular}{ccccccc}
    \vspace{3pt}
    \thead{Quantum channel} & $A$ & $\bm{c}$ & $\mathcal{F}_{\theta}$ & $\mathcal{F}_{\phi}$ & $\mathcal{I}_{\theta}$ & $\mathcal{I}_{\phi}$\\\hline\vspace{6pt}

    \thead{\tabincell{c}{Phase-damping\\channel\\(PDC)}} &
    $\left(\begin{array}{ccc}
    s & 0 & 0\\
    0 & s & 0\\
    0 & 0 & 1
    \end{array}\right)$ & $\left(\begin{array}{c}
    0\\
    0\\
    0
    \end{array}\right)$ & $1$ & $s^{2}\sin^{2}\theta$ &
    $\frac{3+4X+2\,s^{2}\cos^{2}\theta-\cos\left(2\,\theta\right)}{2\left(1+X\right)^{2}}$ &
    $\frac{2s^{2}\sin^{2}\theta}{1+X}$ \\\vspace{6pt}

    \thead{\tabincell{c}{Depolarizing\\channel\\(DPC)}} & $\left(\begin{array}{ccc}
    s & 0 & 0\\
    0 & s & 0\\
    0 & 0 & s
    \end{array}\right)$ & $\left(\begin{array}{c}
    0\\
    0\\
    0
    \end{array}\right)$ & $s^{2}$ & $s^{2}\sin^{2}\theta$ &
    $2-2\sqrt{1-s^{2}}$ & $\frac{2s^{2}\sin^{2}\theta}{1+\sqrt{1-s^{2}}}$ \\\vspace{6pt}

    \thead{\tabincell{c}{Amplitude Damping\\channel\\(ADC)}} & $\left(\begin{array}{ccc}
    \sqrt{s} & 0 & 0\\
    0 & \sqrt{s} & 0\\
    0 & 0 & s
    \end{array}\right)$ & $\left(\begin{array}{c}
    0\\
    0\\
    -p
    \end{array}\right)$ & $s$ & $s\sin^{2}\theta$ &
    $\frac{s\left[3+4Y+2\,s\sin^{2}\theta+\cos\left(2\,\theta\right)\right]}{2\left(1+Y\right)^{2}}$ &
    $\frac{2s\sin^{2}\theta}{1+Y}$\\\vspace{3pt}

    \tabincell{c}{Generalized Amplitude\\Damping channel\\(GADC)} & $\left(\begin{array}{ccc}
    \sqrt{\bar{s}} & 0 & 0\\
    0 & \sqrt{\bar{s}} & 0\\
    0 & 0 & \bar{s}
    \end{array}\right)$ & $\left(\begin{array}{ccc}
    0\\
    0\\
    -\bar{p}\left(\alpha-\beta\right)
    \end{array}\right)$ & $\frac{R_{1}}{2}+\frac{R_{2}}{Z}$ &
    $\bar{s}\sin^{2}\theta$ & $\frac{R_{1}}{1+Z}+R_{2}\left[\frac{1}{Z^{2}}-\frac{1}{\left(1+Z\right)^{2}}\right]$ &
    $\frac{2\bar{s}\sin^{2}\theta}{1+Z}$
  \end{tabular}
  \end{ruledtabular}
\end{table*}

\paragraph{Phase-damping channel:}

The PDC is a prototype model of dephasing or pure decoherence, i.e.,
loss of coherence of a two-level state without any loss of system's energy.
The PDC is described by the map
\begin{equation}
  {\cal E}_{{\rm PDC}}(\rho)=s\,\rho+p\,\left(\rho_{00}|0\rangle\langle0|+\rho_{11}|1\rangle\langle1|\right),\label{eq:PDC}
\end{equation}
and obviously the Kraus operators are given by
\begin{equation}
  \mathbf{K}_{{\rm PDC}}=\left\{ \sqrt{s}\,\openone,\;\sqrt{p}\,|0\rangle\langle0|,\;\sqrt{p}\,|1\rangle\langle1\right\},\label{eq:PDC_kraus}
\end{equation}
where $p\equiv1-s$ is the probability of the qubit exchanging
a quantum with the bath at time $t$ with $s=\exp\left(-\gamma t/2\right)$,
with $\gamma$ denoting the zero-temperature dissipation rate.
For the PDC, there is no energy change and a loss of
decoherence occurs with probability $p$.
As a result of the action of the PDC, the Bloch sphere is
compressed by a factor $(1-2p)$ in the $xy$-plane.

According to Eq.~\eqref{eq:affine_map_qubit},
the transformation matrix $A$ and the translation
vector $\bm{c}$ of the PDC are given in Table \ref{tab:Table1}.
These indicate that the Bloch vector components
along the $x$- and $y$-axis shrink with probability $s$,
while the $z$-component remains invariant under the
action of the PDC, and the Bloch vector $\bm{\omega}$ is mapped as
\begin{equation}
  \mathcal{E}_{{\rm PDC}}\left(\bm{\omega}\right)=\left(s\,\omega_{x},s\,\omega_{y},\omega_{z}\right)^{\rm T}.
\end{equation}
Furthermore, with Eqs.~\eqref{eq:QFI1_noisy} and \eqref{eq:QFI2_noisy},
we obtain the analytical results of the two variant
versions of the QFI under the PDC in Table.~\ref{tab:Table1}.
One can find that $\mathcal{F}_\phi$, $\mathcal{I}_\theta$
and $\mathcal{I}_\phi$ are monotonic functions of $t$
and are solely dependent on the parameter $\theta$.
Interestingly, $\mathcal{F}_\theta$ is constantly
equal to $1$ for any time, which implies that the QFI
$\mathcal{F}$ about the amplitude parameter $\theta$ is robust under PDC.
The significance of this is that one can avoid the
impact of the PDC on the accuracy of the parameter estimation,
by encoding the parameter on the amplitude of the input state.

\begin{figure}[tbp!]
  \centering
  \includegraphics[width=8.5cm,height=9cm]{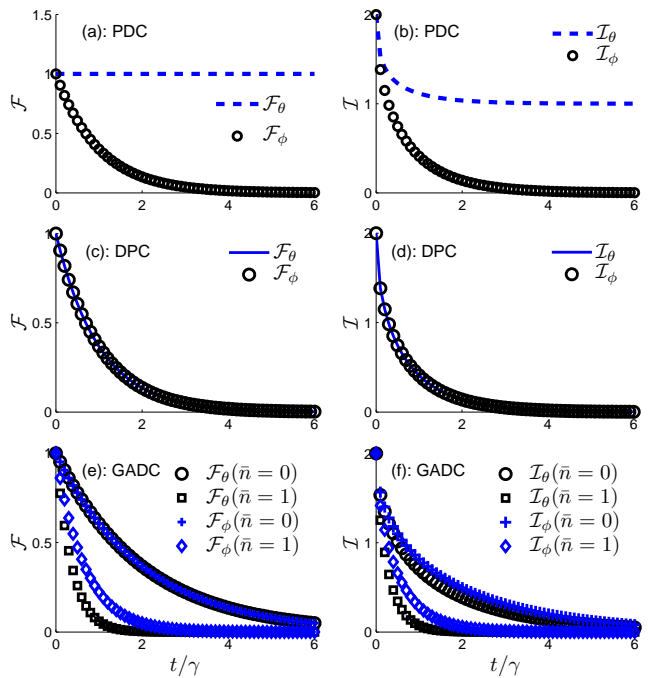}
  \caption{(Color online) Plots of the two variant versions of the QFI
  versus $t$ in terms of the parameter $\theta$ and $\phi$ under three
  quantum decoherence channels: PDC (a,\,b), DPC (c,\,d), and GADC (e,\,f) for a single qubit system.
  The qubit initially being prepared in the equally weighted state
  (Eq.~\eqref{eq:ini_state} with $\theta=\pi/2$) of which the QFIs take the maximum at $t=0$.
  In (e) and (f), we plot the QFIs under the GADC for a zero-temperature
  $\bar{n}=0$ and a finite-temperature $\bar{n}=1$, respectively. \label{fig:QFI_channel}}
\end{figure}

When $\theta=\pi/2$, we have
\begin{eqnarray}
  \mathcal{F}_\theta &=& 1,\ \mathcal{F}_\phi = s^{2},\nonumber\\
  \mathcal{I}_\theta &=& \frac{2}{1+\sqrt{1-s^{2}}},\nonumber\\
  \mathcal{I}_\phi &=& 2-2\sqrt{1-s^{2}}.\nonumber
\end{eqnarray}
As is plotted in Figs.~\ref{fig:QFI_channel}~(a) and (b),
under the PDC, the QFIs associated to the phase parameter,
$\mathcal{F}_\phi$ and $\mathcal{I}_\phi$ decreases
monotonically with $t$ and vanish only in the asymptotic limit $t\rightarrow\infty$.
Interestingly, $\mathcal{I}_\theta$ decrease from
initial value $2$ to final value $1$, shown in Fig.~\ref{fig:QFI_channel}~(b).
It is indicated that the QFI $\mathcal{I}_\theta$
depend on the diagonal and off-diagonal elements of the density matrix.
Due to the influence of the PDC, the off-diagonal ones vanish,
and only the diagonal ones remain, which make $\mathcal{I}_\theta$ equal to $1$.

\paragraph{Depolarizing channel:}

The definition of the DPC is given via the map
\begin{equation}
  {\cal E}_{{\rm DPC}}(\rho) = s\,\rho+p\,\frac{\openone}{2},
  \label{eq:DPC}
\end{equation}
and the corresponding Kraus operators are expressed as
\begin{eqnarray}
  \mathbf{K}_{{\rm DPC}} & = &
  \left\{\frac{\sqrt{1+3s}}{2}\openone,\,\frac{\sqrt{p}}{2}\sigma_{x},\,\frac{\sqrt{p}}{2}\sigma_{y},\,\frac{\sqrt{p}}{2}\sigma_{z}\right\},
  \label{eq:DPC_kraus}
\end{eqnarray}
where $s\equiv1-p$ and $\bm{\sigma}=\left(\sigma_x,\,\sigma_y,\,\sigma_z\right)^{\rm T}$
denotes the Pauli matrices.
The channel is completely positive for $0\leq p\leq1$.
We see that for the DPC, the qubit is unchanged with
probability $s$ or it is depolarized to the completely
mixed state $\openone/2$ with probability $p$.
It is seen that due to the action of the DPC,
the radius of the Bloch sphere is reduced by a factor $s$,
but its shape remains unchanged.

The transformation matrix $A$ and the translation vector
$\bm{c}$ for the DPC are given in Table \ref{tab:Table1},
and then the affine-mapped Bloch vector is obtained as
\begin{equation}
  \mathcal{E}_{{\rm DPC}}\left(\bm{\omega}\right)=
  \left(s\,\omega_{x},\,s\,\omega_{y},\,s\,\omega_{z}\right)^{\rm T},
\end{equation}
which shows that all components of the Bloch
vector are shortened by a factor $s$.
Moreover, we analytically derive the expressions of
the two QFIs with respect to $\theta$ and $\phi$ under the DPC,
given in Table \ref{tab:Table1}.

For $\theta=\pi/2$, those expressions are explicitly simplified as
\begin{eqnarray}
  \mathcal{F}_\theta &=& \mathcal{F}_\phi = s^{2},\nonumber\\
  \mathcal{I}_\theta &=& \mathcal{I}_\phi = 2-2\sqrt{1-s^{2}}.\nonumber
\end{eqnarray}
The results show that, under the DPC, the two QFIs
decrease by a factor of $s^2$ and $s$, respectively.
The expressions of the QFI $\mathcal{F}$ ($\mathcal{I}$)
for different parameters $\theta$ and $\phi$ are the same.
As are plotted in Figs.~\ref{fig:QFI_channel}~(c) and (d),
the QFIs decreases monotonically.

\paragraph{Generalized Amplitude-damping channel:}

The GADC is given, in the Born-Markov approximation,
via its Kraus representation as
\begin{equation}
  {\cal E}_{{\rm GADC}}(\rho)=\sum_{i=0}^{3}K_{i}\,\rho\,K_{i}^{\dagger},
  \label{eq:GADC}
\end{equation}
where the corresponding Kraus operators are
\begin{equation}
  \begin{aligned}
    \mathbf{K}_{{\rm GADC}} = &\Big{\{}\sqrt{\alpha}\left(\left|0\right\rangle\left\langle 0\right|+\sqrt{\bar{s}}\left|1\right\rangle\left\langle 1\right|\right),\,
    \sqrt{\alpha\,\bar{p}}\left|0\right\rangle\left\langle 1\right|,\,\\
    &\sqrt{\beta}\left(\sqrt{\bar{s}}\left|0\right\rangle\left\langle 0\right|+\left|1\right\rangle \left\langle 1\right|\right),\,
    \sqrt{\beta\,\bar{p}}\left|1\right\rangle\left\langle 0\right|\Big{\}},
    \label{eq:GADC_kraus}
  \end{aligned}
\end{equation}
with
\begin{equation}
  \alpha\equiv\frac{\bar{n}+1}{2\,\bar{n}+1},\ \beta\equiv\frac{\bar{n}}{2\,\bar{n}+1},
\end{equation}
and
\begin{equation}
  \bar{s}\equiv e^{-\frac{1}{2}\gamma\,t\left(2\,\bar{n}+1\right)},\ \bar{p}=1-\bar{s},
\end{equation}
where $\bar{s}$ and $\bar{p}$ are dependent on
the mean number of excitations $\bar{n}$ in the bath.
In the zero-temperature limit, i.e., $\bar{n}=0$
and $\bar{s}=s $, Eq.~\eqref{eq:GADC} reduces to
the purely dissipative ADC, and its Kraus
operators are represented as
\begin{eqnarray}
  \mathbf{K}_{{\rm ADC}} & = &
  \left\{ \sqrt{s}\,|0\rangle\langle0|+|1\rangle\langle1|,\,\sqrt{p}\,|1\rangle\langle0|\right\}.
  \label{eq:kraus_ADC}
\end{eqnarray}

Similarly, in the Bloch representation,
the GADC can be described as an affine
map of which the transformation matrix
$A$ and the translation vector $\bm{c}$
being given Table \ref{tab:Table1},
and the Bloch vector $\bm{\omega}$ is
mapped as
\begin{equation}
  \mathcal{E}_{{\rm GADC}}\left(\bm{\omega}\right)=\left(\sqrt{\bar{s}}\,\omega_{x},\sqrt{\bar{s}}\,\omega_{y},\bar{s}\,\omega_{z}-\bar{p}\left(\alpha-\beta\right)\right)^{\rm T}.\label{eq:map_GADC}
\end{equation}
When $\alpha=1$ and $\beta=0$,
Eq.~\eqref{eq:map_GADC} reduce to the ADC case.
It indicates that the GADC squeezes
the Bloch sphere into an ellipsoid
and shifts it towards the north pole.
The radius in the $xy$-plane is reduced
by a factor $\sqrt{s}$, while in the
$z$-direction it is reduced by a factor $s$.
In the asymptotic limit $t\rightarrow\infty$,
i.e., $s=0,\,p=1$, the Bloch vector becomes
$\mathcal{E}_{{\rm GADC}}\left(\bm{\omega}\right)=\left(0,\,0,\,-\left(\alpha-\beta\right)\right)^{\rm T}$,
which also implies that, under the ADC ($\bar{n}=0$),
the qubit finally stay in the ground state.
Meanwhile, the analytical results of the two
QFIs under the GADC (the finite temperature)
and the ADC (the zero-temperature) are derived in Table \ref{tab:Table1}.

When $\theta=\pi/2$, the dynamics of the QFI
$\mathcal{F}$, under the ADC, in terms of $\theta$
and $\phi$ are the same, namely $\mathcal{F}_{\theta}=\mathcal{F}_{\phi}=s$.
As are shown in Figs.~\ref{fig:QFI_channel}~(e) and (f),
under the GADC and ADC, the QFIs with for the different
parameters decrease monotonically with time.
It is also shown that the QFIs for finite temperature
decay more rapidly than that for zero-temperature.

\subsection{Numerical calculation with hierarchy equation}\label{num}

In this section, we focus on a simple dissipative model of a two-level system
interacting with a zero-temperature bosonic reservoir to explicitly illustrate
the behaviors of the two QFI quantities during time evolution \cite{BreuerBook,Lu2010}.
Here, we exactly examine this model by adopting the hierarchy equation method.
The total Hamiltonian of the system and bath without performing the RWA is
\begin{equation}
  H = \frac{1}{2}\omega_{0}\sigma_{z}+\sum_{k}\omega_{k}b_{k}^{\dagger}b_{k}+\sigma_{x}B,\label{eq:Ham_qubit}
\end{equation}
with $B=\sum_{k}g_{k}b_{k}+{\rm h.c.}$.
Here, the first term is the free Hamiltonian of the qubit with
transition frequency $\omega_{0}$, the second term denotes the
environment part with the creation (annihilation) operators $b_{k}^{\dagger} \left(b_{k}\right)$
of the bath model with frequency $\omega_{k}$, and the last term
is the interaction Hamiltonian between the system and bath
equipped with the coupling constant $g_{k}$.
In the zero temperature limit, the spectral density is generally
represented by a Lorentzian \cite{BreuerBook,Gorini1976}
\begin{equation}
  J\left(\omega\right) = \frac{1}{\pi}\frac{\lambda\gamma}{\left(\omega-\omega_{0}\right)^{2}+\gamma^{2}},\label{eq:Loren}
\end{equation}
where $\lambda$ reflects the system-bath coupling strength
and $\gamma$ is the spectral width of the coupling,
which is related to the reservoir correlation time scale $\tau_{B}\sim\gamma^{-1}$.

As is well known, this typical dissipative model is solvable
under the RWA which is effective in the weak coupling limit \cite{BreuerBook,Ma2012}.
With the RWA, the analytical time-evolution function of the system
can be equivalently described as an ADC by redefining the
Kraus operators Eq.~\eqref{eq:kraus_ADC} as
\begin{equation}
  \widetilde{\mathbf{K}}_{{\rm ADC}}=
  \left\{h\left(t\right)\,|0\rangle\langle0|+|1\rangle\langle1|,\,\sqrt{\left(1-h^2\left(t\right)\right)}\,|1\rangle\langle0|\right\} \label{eq:kraus_JC}
\end{equation}
where $h\left(t\right)$ is a crucial characteristic function as
\begin{equation}
  h\left(t\right)=
  e^{-\gamma t/2}\left[\cosh\left(\frac{dt}{2}\right)+\frac{\gamma}{d}\sinh\left(\frac{dt}{2}\right)\right],
\end{equation}
with $d = \sqrt{\gamma^{2}-4\lambda}$.
From Eq.~\eqref{eq:map_GADC}, the affine-mapped bloch vector
reads under this dissipative environment
\begin{equation}
    \widetilde{\mathcal{E}}_{{\rm ADC}}\left(\bm{\omega}\right)=
    \left(h\left(t\right)\,\omega_{x},\,h\left(t\right)\,\omega_{y},\,1-h^2\left(t\right)\right)^{\rm T}.
\end{equation}
As shown in Table \ref{tab:Table1},
then, one can easily recover the result of the dynamical QFI
in terms of $\phi$ given in Ref.~\cite{Lu2010},
\begin{equation}
  \mathcal{F}_{\phi} = h^{2}\!\left(t\right).\label{eq:QFI_lu}
\end{equation}

With the hierarchy equation method, the exact
dynamic of the system is derived as the following
equation in the interaction picture \cite{Ma2012}
\begin{eqnarray}
    \rho_{{\rm S}}^{\left(I\right)}\left(t\right)
    & = & \mathcal{T}\exp\Big{\{}-\int_{0}^{t}dt_{2}\int_{0}^{t_{2}}dt_{1}V\left(t_{2}\right)^{\times}\Big{[}C^{R}\left(t_{2}-t_{1}\right)
    \nonumber\\
    & & V\left(t_{1}\right)^{\times}+iC^{I}\left(t_{2}-t_{1}\right)V\left(t_{1}\right)^{\circ}\Big{]}\Big{\}}\rho_{{\rm S}\left(0\right)},
    \label{eq:HierarchyEq}
\end{eqnarray}
where $\mathcal{T}$ is the chronological time-ordering operator and,
to simplify the description, we introduce two super-operators
$A^{\times}B\equiv\left[A,B\right]$ and $A^{\circ}B\equiv\left\{A,B\right\}$.
Assume that the initial state is taken as $\rho\left(0\right)=\rho_{{\rm S}}\left(0\right)\,\otimes\,\rho_{{\rm B}}$
with $\rho_{{\rm B}}$ being in the vacuum state $\otimes_{k}\left|0_{k}\right\rangle$.
Also, $C^{R}\left(t_{2}-t_{1}\right)$ and $C^{I}\left(t_{2}-t_{1}\right)$
respectively correspond to the real and imaginary part of the
bath time-correlation function which is defined as
\begin{equation}
  C\left(t_{2}-t_{1}\right)\equiv\left\langle B\left(t_{2}\right)B\left(t_{1}\right)\right\rangle_{{\rm B}}
  = {\rm Tr}\left[B\left(t_{2}\right)B\left(t_{1}\right)\rho_{{\rm B}}\right],
  \label{eq:timecorrelation}
\end{equation}
where $B\left(t\right) = \sum_{k}g_{k}b_{k}e^{-i\omega_{k}t}+{\rm h.c.}$.
Then the time-correlation function Eq.~\eqref{eq:timecorrelation}
becomes the exponential form
\begin{equation}
  C\left(t_{2}-t_{1}\right) =
  \lambda\exp\left[-\left(\gamma+i\omega_{0}\right)\left|t_{2}-t_{1}\right|\right],
  \label{eq:timecorrelation_analytical}
\end{equation}
With Eqs.~\eqref{eq:HierarchyEq} and \eqref{eq:timecorrelation_analytical},
we further obtain the set of hierarchical equations
of the qubit as \cite{Ma2012}
\begin{eqnarray}
  \frac{\partial}{\partial t}\varrho_{\vec{n}}\left(t\right)
  & = & -\left(iH_{S}^{\times}+\vec{n}\cdot\vec{\nu}\right)\varrho_{\vec{n}}\left(t\right)
  -i\sum_{k=1}^{2}V^{\times}\varrho_{\vec{n}+\vec{e}_{k}}\left(t\right)\nonumber\\
  & - & i\frac{\lambda}{2}\sum_{k=1}^{2}n_{k}\left[V^{\times}+
  \left(-1\right)^{k}V^{\circ}\right]\varrho_{\vec{n}-\vec{e}_{k}}\left(t\right),
  \label{eq:hier_eq}
\end{eqnarray}
where the subscript $\vec{n}=\left(n_{1},\, n_{2}\right)$ is
a two-dimensional index with $n_{1\left(2\right)}\ge0$,
and $\rho_{S}\left(t\right)\equiv\varrho_{\left(0,0\right)}\left(t\right)$.
The vectors are $\vec{e}_{1}=\left(1,\,0\right)$, $\vec{e}_{2}=\left(0,\,1\right)$,
and $\vec{\nu}=\left(\nu_{1},\,\nu_{2}\right)=\left(\gamma-i\omega_{0},\,\gamma+i\omega_{0}\right)$.
We emphasize that $\varrho_{\vec{n}}\left(t\right)$, with $\vec{n}\ne\left(0,\,0\right)$,
are auxiliary operators introduced only for the sake of computing,
they are not density matrices, and are all set to be zero at $t=0$.
Through solving the above hierarchy equations, the dynamics of
the system can be exactly determined without making the RWA.

\begin{figure}[h]
  \begin{centering}
  \includegraphics[width=8.8cm,height=6.5cm]{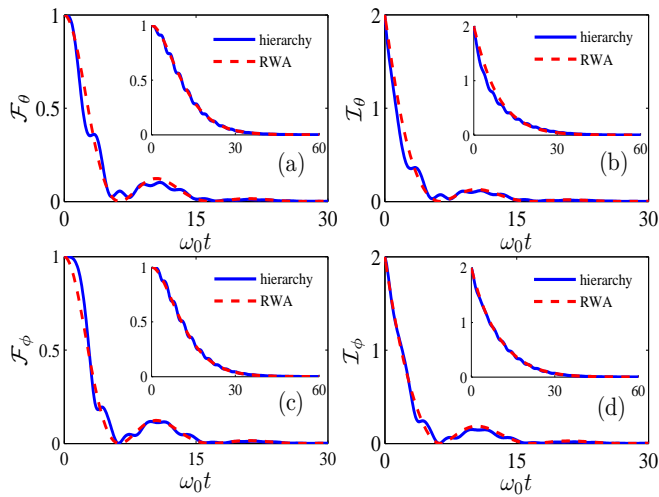}
  \par\end{centering}
  \caption{(Color online) Two QFIs Eq.~\eqref{eq:QFI1}
  (a,\,b) and Eq.~\eqref{eq:QFI2} (c,\,d) versus time
  for different parameters initially embedded in the qubit under the dissipative process.
  The solid lines display the numerical results by using the hierarchy equation method.
  These are compared to the analytical results with RWA (dashed lines),
  all plotted in the non-Markovian regime ($\lambda=0.1\gamma$).
  The insets show in the Markovian regime ($\lambda=0.01\gamma$)}\label{fig:QFIs}
\end{figure}

Figure~\ref{fig:QFIs} displays the dynamics of the QFI of
Eq.~\eqref{eq:QFI1} and the variant QFI of Eq.~\eqref{eq:QFI2}
versus time for different parameters initially encoded in
the qubit state. The spectral width of the coupling is set
by $\gamma=0.2\omega_{0}$. In the Markovian regime (inset plots with $\lambda=0.01\gamma$),
the QFIs $\mathcal{F}_{\theta\left(\phi\right)}$ and $\mathcal{I}_{\theta\left(\phi\right)}$
are monotonically go to zero. Both the numerical and the
analytical results are consistent in the weak coupling limit.
While, in the non-Makovian regime ($\gamma=0.1\lambda$),
due to the strong coupling with the reservoir, the dynamics
of these two information quantities exhibit the oscillations
and revivals over time. The times when $\mathcal{I}_{\theta\left(\phi\right)}$
vanish completely coincide with those for $\mathcal{F}_{\theta\left(\phi\right)}$.
The surprising aspect here is that the quantities of
$\mathcal{I}_{\theta\left(\phi\right)}$ for the two parameters
$\theta$ and $\phi$ decrease quickly to zero from the initial
value of $2$, and then later on revive with almost the same
amount as $\mathcal{F}_{\theta\left(\phi\right)}$ do.
Since the hierarchical equations we obtained exactly depict
the dynamics of the system, the oscillating deviations between
the numerical and the analytical results can be interpreted as
the contribution of the counter-rotating-wave terms which have
been omitted by making the RWA \cite{Ma2012}.
Such deviations behave evidently for both two information
quantities in terms of $\theta$. As shown in Ref.~\cite{Lu2010},
they addressed these oscillations and revivals in Figs.~\ref{fig:QFIs}~(a)
and (b) by introducing the definition of the QFI flow.
Analogously, we also introduce a variant QFI flow (or skew information flow)
as the change rate of information quantity $\mathcal{I}$, i.e.,
$\sigma:=\partial\mathcal{I}_{\lambda}/\partial t$.
Also, $\sigma>0$ indicates that there is a information flow from
the environment to the system, corresponding to the revivals, and
$\sigma<0$ denotes that the flow from the system to the environment,
accounting for the decays, as shown in Figs.~\ref{fig:QFIs}~(c) and (d).

\section{Generalization}\label{QFI_qudit}

\subsection{Representation of the QFI in terms of the generalized Bloch vector for a qudit system}

We have examined the QFIs under decoherence for
a single qubit in the Bloch representation.
Now, we consider a more general case, i.e.,
the qudit (a $d$-dimensional quantum system).
A general qudit state can be written in the Bloch
representation as \cite{Lendi1987,LendiBook2007}
\begin{equation}
  \rho = \frac{1}{d}\,\openone_{d}+\frac{1}{2}\,\bm{\omega}\cdot\hat{\bm{\eta}},
  \label{eq:Bloch_rep}
\end{equation}
where $\openone_{d}$ is a $d\times d$ identity operator,
$\hat{\bm{\eta}}={\left\{\hat{\eta_{i}}\right\}}_{i=1}^{d^2-1}$
are the generators of the Lie algebra $\mathfrak{su}\left(d\right)$
(see appendix \ref{sec:AppI}), and $\bm{\omega}\in\mathbb{R}^{d^2-1}$
denotes the generalized Bloch vector whose $i$th element
is ${\rm Tr}\left(\rho\,\hat{\eta}_{i}\right)$.

Since we have the following relation
\begin{equation}
  1/d\leq{\rm Tr}\left(\rho^{2}\right)\leq1,
  \label{eq:purity_relation}
\end{equation}
where ${\rm Tr}\left(\rho^{2}\right)$ is the purity.
The purity equal to one, corresponding to pure states,
and $1/d$, corresponding to mixed states.
With Eq.~\eqref{eq:Bloch_rep}, we get
\begin{equation}
  {\rm Tr}\left(\rho^{2}\right)=\frac{1}{d}+\frac{1}{2}\left|\bm{\omega}\right|^{2},
  \label{eq:purity}
\end{equation}
by using following relation
\begin{equation}
  {\rm Tr}\left[\left(\bm{a}\cdot\hat{\bm{\eta}}\right)\left(\bm{b}\cdot\hat{\bm{\eta}}\right)\right]
  =2\,\bm{a}\cdot\bm{b}.\label{eq:qudit_Tr_relation}
\end{equation}
Thus, from Eqs.~\eqref{eq:purity_relation} and \eqref{eq:purity},
we obtain the length of the generalized Bloch vector satisfying
\begin{equation}
  0\leq\omega\leq\sqrt{2\left(d-1\right)/d},
\end{equation}
where the first equality holds for the maximal mixed state
and the second for pure states.

Owning to the completeness of the generators of Lie algebra,
any Hermitian matrix can be described by the common set of generators.
Based on this representation, we give the new expressions
of the two QFIs in terms of the Bloch vector.
Firstly, we consider the QFI in Eq.~\eqref{eq:QFI1}.
$\mathcal{F}_{\lambda}$ can be represented as
\begin{equation}
  \mathcal{F}_{\lambda}=
  \begin{cases}
  \left(\partial_\lambda\bm{\omega}\right)^{\rm T}\,\mathcal{M}^{-1}\,\partial_{\lambda}\bm{\omega}, & \omega<\sqrt{2\left(d-1\right)/d},\\
  \left|\partial_{\lambda}\bm{\omega}\right|^{2}, & \omega=\sqrt{2\left(d-1\right)/d},\label{eq:QFI1_qudit}
\end{cases}
\end{equation}
where $\mathcal{M}$ is a real symmetry matrix defined as
\begin{equation}
  \mathcal{M}=\frac{2}{d}\openone_{d^2-1}-\bm{\omega}\bm{\omega}^{\rm T}+G.\label{eq:M_matrix}
\end{equation}
The superscript ${\rm T}$ in the above equations denotes the transpose
operation and $\mathcal{M}^{-1}$ denotes the matrix inverse of $\mathcal{M}$.
$\openone_{d^2-1}$ is the identity matrix of dimension $d^2-1$ and $G$
is a $\left(d^2-1\right)\times\left(d^2-1\right)$ real symmetric matrix
whose $ij$-element is
\begin{equation}
  \left[\,G\,\right]_{ij}=\sum_{k=1}^{d^2-1}\,g_{ijk}\,\omega_k, \label{eq:G_matrix}
\end{equation}
where $g_{ijk}$ is the completely symmetric tensor defined in Eq.~\eqref{eq:tensor_sym}.
Hence, $\mathcal{M}$ is also real symmetric matrix.
Since $\mathcal{M}$ may have some zero eigenvalues,
the inverse is defined on the support of $\mathcal{M}$ \cite{Ueda2010}.
Here, the first line of Eq.~\eqref{eq:QFI1_qudit} only
applies to mixed states, and the detailed derivation
can be found in appendix~B.

For pure states, $\mathcal{F}_{\lambda}$ is generally expressed
as the norm of the derivative of the Bloch vector shown in Eq.~\eqref{eq:QFI1_qudit}.
It can be easily derived by using Eq.~\eqref{eq:qudit_Tr_relation}
and following the same procedure used in deriving the second line of Eq.~\eqref{eq:QFI1_qubit}.

From Eq.~\eqref{eq:M_matrix}, we can see that the matrix $\mathcal{M}$
is dependent of $\bm{\omega}$. Then we can conclude that $\mathcal{F}_{\lambda}$
given by Eq.~\eqref{eq:QFI1_qudit} only depends on the two elements: the Bloch
vector of the density matrix and the derivative of it. Moreover,
it deserves to emphasize that Eq.~\eqref{eq:QFI1_qudit} is also valid
for arbitrary quantum state, since a density matrix always can be
expressed as the form of Eq.~\eqref{eq:Bloch_rep} by expanding over the
generators of the Lie algebra.

When $d=2$, we have $G=0$, due to $g_{ijk}=0$.
Then, the real symmetric matrix $\mathcal{M}$ reduces to
\begin{equation}
  \mathcal{M}=\openone_{3}-\bm{\omega}\bm{\omega}^{\rm T}.
\end{equation}
The inverse of $\mathcal{M}$ is verified as
\begin{equation}
  \mathcal{M}^{-1}=\openone_{3}+\frac{\bm{\omega}\,\bm{\omega}^{\rm T}}{1-\left|\bm{\omega}\right|^{2}}.
\end{equation}
Substituting the above equation into Eq.~\eqref{eq:QFI1_qudit}
finally recovers the first line of Eq.~\eqref{eq:QFI1_qubit}.

We next take account of the QFI in Eq.~\eqref{eq:QFI2}.
We firstly expand
\begin{equation}
  \sqrt{\rho}=y\openone_{d}+\bm{x}\cdot\hat{\bm{\eta}}.\label{eq:sqrt_rho}
\end{equation}
Then, $\mathcal{I}_{\lambda}$ can be represented as follows
\begin{equation}
  \mathcal{I}_{\lambda}=
  \begin{cases}
  8\left[\left(\partial_{\lambda}y\right)^2+\frac{2}{d}\left|\partial_{\lambda}\bm{x}\right|^2\right], & \omega<\sqrt{2\left(d-1\right)/d},\\
  2\left|\partial_{\lambda}\bm{\omega}\right|^{2}, & \omega=\sqrt{2\left(d-1\right)/d},\label{eq:QFI2_qudit}
\end{cases}
\end{equation}
where $y$ and $\bm{x}$ are completely determined by the following
$d^2$ quadratic nonlinear equations:
\begin{eqnarray}
  y^2+\frac{2}{d}\left|\bm{x}\right|^2 &=& \frac{1}{d},\label{eq:coef_y}\\
  2\,y\,x_{k}+\sum_{i,j=1}^{d^2-1}g_{ijk}\,x_{i}\,x_{j} &=& \frac{\omega_{k}}{2},\label{eq:vec_x}
\end{eqnarray}
for $k=1,2,...,d^2-1$.
The first line of the above equation is applicable for mixed states,
and the detailed derivation is given in appendix~C.

For pure states, $\mathcal{I}_{\lambda}$ is generally expressed as
the norm of the derivative of the Bloch vector up to a factor of $2$
shown in Eq.~\eqref{eq:QFI2_qudit}.
It can be obtained by using Eq.~\eqref{eq:qudit_Tr_relation}
and following the same procedure used in deriving the second line of Eq.~\eqref{eq:QFI2_qubit}.

For the case of $d=2$, equations~\eqref{eq:coef_y}
and \eqref{eq:vec_x} reduce to
\begin{eqnarray}
  y^2+\left|\bm{x}\right|^2 &=& \frac{1}{2}, \label{eq:xy_fun1}\\
  4\,y\,\bm{x} &=& \bm{\omega}, \label{eq:xy_fun2}
\end{eqnarray}
with $g_{ijk}=0$. By solving the above equations (see appendix~C),
one can obtain
\begin{eqnarray}
  y &=& \frac{\sqrt{1+\sqrt{1-\left|\bm{\omega}\right|^2}}}{2}, \label{eq:y_sol}\\
  \bm{x} &=& \frac{\bm{\omega}}{2\sqrt{1+\sqrt{1-\left|\bm{\omega}\right|^2}}}. \label{eq:x_sol}
\end{eqnarray}
Inserting the above solutions into the first line of Eq.~\eqref{eq:QFI2_qudit}
and making some simplification recovers the first line
given in Eq.~\eqref{eq:QFI2_qubit}.

\subsection{QFI for an $N$-qubit system in noisy environment}

Below, we study the dynamic of the QFI for an $N$-qubit system in noisy environment.
In the Bloch representation, the two QFIs for qudit are described
in terms of the generalized Bloch vector $\bm{\omega}$,
as is shown in Eqs.~\eqref{eq:QFI1_qudit} and ~\eqref{eq:QFI2_qudit} respectively.
The Bloch vector is assumed to be the function of an unknown parameter $\lambda$ on the system.
Meanwhile, we emphasize that Eqs.~\eqref{eq:QFI1_qudit} and ~\eqref{eq:QFI2_qudit}
are also applicable for the multi-qubit system with exchange symmetry.
Since, a collection of $N$ qubits is represented by the collective operators \cite{Wang2003}
\begin{equation}
  J_{\alpha}=\sum_{i=1}^{N}\frac{\sigma_{i\alpha}}{2},\,\left(\alpha=x,\ y,\ z\right),
\end{equation}
where $\sigma_{i\alpha}$ denotes the pauli matrix of the $i$th qubit.
Such an $N$-qubit ensemble with total angular momentum $j=N/2$ can be approximately viewed as a qudit system,
when it has the symmetry under the exchange of two qubits.
The collective basis of this system is $\left\{\left| j,m\right\rangle\right\}$ for $m=0,\pm1,\pm2,...,\pm j$,
which is so-called Dicke state, $J_z\left| j,m\right\rangle=m\left| j,m\right\rangle$.
Hence, Eqs.~\eqref{eq:QFI1_qudit} and ~\eqref{eq:QFI2_qudit} are also valid for those multi-qubit systems.

Assume that the dimension of the decohered state
$\mathcal{E}(\rho)$ is the same of that of $\rho$.
Similar to Eq.~\eqref{eq:affine_map_qubit},
an $N$-qubit system with exchange symmetry described by
Eq.~\eqref{eq:Bloch_rep} under decoherence can be expressed as
\cite{Ueda2010,NielsenBook}
\begin{equation}
  \mathcal{E}(\rho)=\frac{1}{d}\openone_{d}+\frac{1}{2}\left(A\,\bm{\omega}+\bm{{\rm c}}\right)\cdot\hat{\bm{\eta}},
  \label{eq:affine_map_qudit}
\end{equation}
with $d=N+1$.
Here, $A$ is a matrix of dimension $d^2-1$ with elements
\begin{equation}
  A_{ij}=\frac{1}{2}{\rm Tr}\left[\hat{\eta}_i\,\mathcal{E}(\,\hat{\eta}_j)\right],
\end{equation}
and $\bm{c}$ is a $d^2-1$ dimensional vector with elements
\begin{equation}
  c_{i}=\frac{1}{d}{\rm Tr}\left[\hat{\eta}_i\,\mathcal{E}(\,\openone_{d})\right].
\end{equation}
Equation~\eqref{eq:affine_map_qudit} illustrates that
a Markovian quantum dynamic can be geometrically described
as an affine transformation, i.e.,
\begin{equation}
  \mathcal{E}:\,\bm{\omega}\mapsto A\,\bm{\omega}+\bm{c}.\label{eq:affine_map_vec}
\end{equation}
As we mentioned at the beginning of Sec.~III, in the parameter estimation,
the unknown parameter is generally imprinted into the probes through unitary
or non-unitary evolution \cite{Escher2011}. It is noted that equation~\eqref{eq:QFI1_qudit}
is applicable to the cases with different the parametrization processes,
by replacing the Bloch vector $\bm{\omega}$ with the affine-mapped Bloch
vector $\mathcal{E}(\bm{\omega})$.

\begin{figure}[tfb]
  \centering
  \includegraphics[width=8cm,height=5cm]{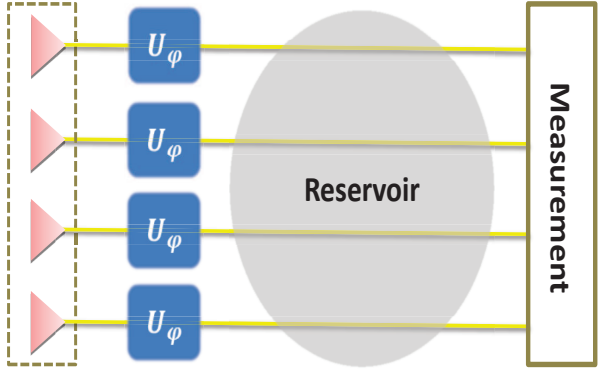}
  \caption{(Color online) The Schematic representation of Ramsey interferometry in presence of collisional dephasing.
  \label{fig:Ramsey}}
\end{figure}

To be more specific, we consider an experimentally realizable
Ramsey interferometry \cite{Boixo2009}, to estimate a physical
parameter in a Bose-Einstein condensate (BEC) of $N$ two-level
atoms interacting with a common thermal reservoir.
This experiment we model is shown schematically in Fig.~\ref{fig:Ramsey}.
Those two-level atoms may be considered as qubits or probes,
which are prepared in an $N$-qubit Greenberger-Horne-Zeilinger (GHZ)
state (or Schr\"odinger-cat state)
\begin{equation}
  \left|\psi_{\rm GHZ}\right\rangle=
  \frac{1}{\sqrt{2}}\left(\left|0\right\rangle^{\otimes N}+\left|1\right\rangle^{\otimes N}\right).
\end{equation}
Considering all the qubits being initially prepared in $\left|0\right\rangle$,
then a GHZ state is generated by putting a Hadamard gate acts on the first qubit,
followed by a sequence of controlled-NOT gates linking the first one
with each of the remaining ones \cite{Huegla1997,Giovannetti2006,Boixo2009}.
It is well known that such a state saturates the ultimate
Heisenberg limit (HL) $1/N$ on precision of measurement.
This result gives a quadratic improvement over the standard
quantum limit (SQL), which is achievable with product states
\cite{Giovannetti2006,Boixo2009}.

As is shown in Fig.~\ref{fig:Ramsey},
the parameter $\varphi$ of interest is unitary imprinted
on the state of the probe qubits.
The unitary operator is given by
\begin{equation}
  U_{\varphi}^{\rm tot}=U_{\varphi}^{\otimes N}=
  \left[\exp\left(-i\frac{\varphi}{2}\sigma_z\right)\right]^{\otimes N}= e^{-i\varphi J_z},
\end{equation}
where $J_{z}$ is the $z$ component of the total angular momentum for all qubits \cite{Boixo2009}.
After the unitary evolution, then the state of the probes becomes
\begin{eqnarray}
  \left|\widetilde\psi_{\rm GHZ}\right\rangle
  &=& \frac{1}{\sqrt{2}}\big{(}\left|0\right\rangle^{\otimes N}+e^{i\,N\varphi}\left|1\right\rangle^{\otimes N}\big{)}\nonumber\\
  &=& \frac{1}{\sqrt{2}}\left(\left|\frac{N}{2},\frac{N}{2}\right\rangle+e^{i\,N\varphi}\left|\frac{N}{2},-\frac{N}{2}\right\rangle\right),
  \label{eq:probes_state}
\end{eqnarray}
up to a global phase factor $e^{-iN\varphi/2}$.
The second equality above is valid in the standard
representation of the generator $J_{z}$.
According to the appendix~D, the Bloch vector of
the state of Eq.~\eqref{eq:probes_state} reads
\begin{eqnarray}
\bm{\omega}_{\varphi} & = & \Big{(}\underset{\left(d^{2}-d\right)/2}{\underbrace{\cos\left(N\varphi\right),\,0,\,...,\,0}},\,
\underset{\left(d^{2}-d\right)/2}{\underbrace{\sin\left(N\varphi\right),\,0,\,...,\,0}},\nonumber\\
 &  & \frac{1}{2},...,\frac{1}{\sqrt{2m\left(m+1\right)}},...,\frac{1}{\sqrt{2\left(d-1\right)\left(d-2\right)}},\nonumber\\
 &  & \frac{2-d}{\sqrt{2d\left(d-1\right)}}\Big{)}^{\rm T},\label{eq:Bloch_vec}
\end{eqnarray}
for $m =1,...,d-2$.
For the sake of clarity, we take $N=2$ (i.e., $d=3$) for example.
The Bloch vector for $N=2$ in Eq.~\eqref{eq:Bloch_vec} becomes
\begin{equation}
  \bm{\omega}_{\varphi}=\big{(}\cos\left(2\varphi\right),\,0,\,0,\,
  \sin\left(2\varphi\right),\,0,\,0,\,\frac{1}{2},\,-\frac{1}{2\sqrt{3}}\big{)}
  ^{\rm T},\nonumber
\end{equation}
of dimension $8$.
As is shown in Eq.~\eqref{eq:Bloch_vec}, $\bm{\omega}_{\varphi}$
only contains two $\varphi$-dependent elements.
From Eq.~\eqref{eq:QFI1_qudit}, we find that that those
$\varphi$-independent elements of $\bm{\omega}_{\varphi}$
will not contribute to computation of $\mathcal{F}_{\varphi}$.

In the realistic experiment, decoherence always exists.
As is shown in Fig.~\ref{fig:Ramsey}, we consider the effect
of the collisional dephasing on this measurement protocol,
which is induced by the interaction between the qubits and
the common thermal reservoir \cite{Jin2010,Gill2011}.
The master equation of the system in the Lindblad form
can be described as \cite{Jin2010,Vardi2009}
\begin{equation}
  \dot{\rho}\left( t\right)
  =\mathcal{L}\,\rho\equiv\gamma\left(2\hat{J}_{z}\rho\left( t\right)\hat{J}_{z}
  -\rho\left( t\right)\hat{J}^2_{z}-\hat{J}^2_{z}\rho\left( t\right)\right),
  \label{eq:master_eq}
\end{equation}
where $\mathcal{L}$ denotes the Lindblad superoperator,
$\gamma$ denotes the dephasing rate, and $\rho$ is the
reduced density operator of the system in the interaction picture.
For the single qubit case, Eq.~\eqref{eq:master_eq} reduces to
\begin{equation}
  \dot{\rho}\left( t\right)
  =\mathcal{L}\,\rho\equiv\frac{\gamma}{2}\left[\sigma_z\rho\left( t\right)\sigma_z-\rho\left( t\right)\right],
  \label{eq:master_eq_qubit}
\end{equation}
which corresponds to a single-qubit dephasing channel (i.e., DPC).
Here, we use the interaction representation which does not
affect the result of the calculation, since the QFI $\mathcal{F}$
remains invariant under the unitary evolution being independent
of the parameter $\varphi$.
From Eq.~\eqref{eq:master_eq}, the time evolution of the
density matrix elements is given as follows
\begin{equation}
  \rho_{m,n}\left(t\right)=
  \left\langle j,m\right|\rho\left(t\right)\left|j,n\right\rangle=
  \rho_{m,n}\left(0\right)e^{-\left(m-n\right)^{2}\gamma t}.
  \label{eq:master_eq result}
\end{equation}
In the Bloch representation, the corresponding affine
transformation matrix for the collisional dephasing in Eq.~\eqref{eq:master_eq}
can be obtained as
\begin{widetext}
  \begin{eqnarray}
  A &=& {\rm Diag}\Big{(}
  \underset{\left(d^{2}-d\right)/2}
  {\underbrace{e^{-N^{2}\gamma t},e^{-\left(N-1\right)^{2}\gamma t},
  e^{-\left(N-1\right)^{2}\gamma t},e^{-\left(N-2\right)^{2}\gamma t},...,e^{-4\gamma t},
  \overset{d-1}{\overbrace{e^{-\gamma t},...,e^{-\gamma t}}}}}\nonumber\\
  & &
  \underset{\left(d^{2}-d\right)/2}
  {\underbrace{e^{-N^{2}\gamma t},e^{-\left(N-1\right)^{2}\gamma t},
  e^{-\left(N-1\right)^{2}\gamma t},e^{-\left(N-2\right)^{2}\gamma t},...,e^{-4\gamma t},
  \overset{d-1}{\overbrace{e^{-\gamma t},...,e^{-\gamma t}}}}},\,
  \underset{d-1}{\underbrace{1,\,1,\,...,\,1}}\Big{)},
  \label{eq:A_matrix}
\end{eqnarray}
\end{widetext}
and $\bm{c}=\bm{0}$ (see appendix D).
Apparently, both transformation matrix $A$ and translation vector
$\bm{c}$ are independent of $\varphi$. After the dephasing process,
the Bloch vector of Eq.~$\eqref{eq:Bloch_vec}$ is affine-mapped as
\begin{eqnarray}
\mathcal{E}\left(\bm{\omega}_{\varphi}\right)
 & = & \Big{(}e^{-N^{2}\gamma t}\cos\left(N\varphi\right),\,0,\,...,\,0,\, e^{-N^{2}\gamma t}\sin\left(N\varphi\right),\nonumber\\
 &  & 0,\,...,\,0,\,\frac{1}{2},...,\,\frac{1}{\sqrt{2k\left(k+1\right)}},...,\nonumber\\
 &  & \frac{1}{\sqrt{2\left(d-1\right)\left(d-2\right)}},\,\frac{2-d}{\sqrt{2d\left(d-1\right)}}\Big{)}^{\rm T},\label{eq:Bloch_vec_dec}
\end{eqnarray}
which is given by inserting $A$ of Eq.~\eqref{eq:A_matrix} and $\bm{c}=\bm{0}$
into Eq.~\eqref{eq:affine_map_vec}.

The dynamic of the QFI associated to the parameter
$\varphi$ may be evaluated as,
from Eqs.~\eqref{eq:QFI1_qudit} and ~\eqref{eq:Bloch_vec_dec},
\begin{equation}
  \mathcal{F}_{\varphi}=N^{2}e^{-2N^2\gamma t},\label{eq:F_col}
\end{equation}
which shows that the quantity of the QFI is monotonically
decreased by a factor of $e^{-2N^2\gamma t}$.
The ultimate accuracy of the estimation reads
\begin{equation}
   \Delta\varphi=\frac{1}{N\,e^{-N^2\gamma t}},
\end{equation}
by inserting Eq.~\eqref{eq:F_col} into Eq.~\eqref{eq:QCRbound}.
It indicates that the collisional dephasing dramatically
deteriorate the sensitivity of the phase $\varphi$ estimation.
Our result is consistent with Ref.~\cite{Dorner2012}.

\begin{figure}[tbf!]
  \centering
  \includegraphics[width=8cm,height=5.5cm]{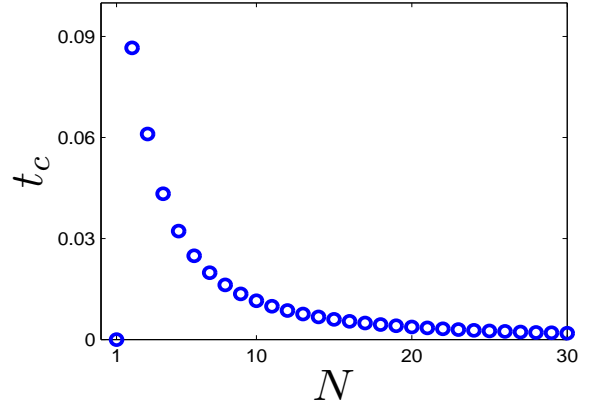}
  \caption{(Color online) Plot of the characteristic time $t_{\bm c}$
  in Eq.~\eqref{eq:t_col} (blue dash-circle curve).\label{fig:QFI_Nqubits}}
\end{figure}

To clearly see the effect of the decoherence process,
we define a time scale $t_{\rm c}$ which is the time over
which the QFI reduce from the HL (i.e., $\mathcal{F}_{\varphi}=N^{2}$)
to the SQL (i.e., $\mathcal{F}_{\varphi}=N$) \cite{Ma2011}.
For the collisional dephasing, the characteristic time reads
\begin{equation}
  t_{\rm c}=\frac{\log N}{2N^{2}},\label{eq:t_col}
\end{equation}
by setting $\gamma=1$.
After $t_{\rm c}$ above, the GHZ state cannot be used to
perform over shot-noise estimation.
As is plotted in Fig.~\ref{fig:QFI_Nqubits},
the characteristic time decreases exponentially as $N$ increases.
It illustrates that the advantage of a GHZ state deteriorates
in the case of collisional dephasing.

\section{Discussion and Conclusion}\label{conclusion}

We have discussed the time evolution of the two variant versions of the QFIs in the presence of quantum noises.
With the help of the Bloch representation, we derived the explicit formula of the two information quantities for a single qubit system.
The analytical expressions of the dynamics of the two QFIs under three typical quantum decoherence channels were obtained.
Both information quantities in those channels are decreased monotonically with time,
only except the case that the $\mathcal{F}_\theta$ remained invariant under PDC.
It manifested that the QFI defined in Eq.~\eqref{eq:QFI1} about the amplitude parameter $\theta$ is robust for the PDC.

We also considered a simple dissipative model of a single qubit coupling with a bosonic reservoir at zero temperature.
By applying the hierarchy equation method, we exactly calculated the dynamical QFIs during time evolution.
We found that the numerical results qualitatively coincided with the analytical ones using the RWA.
The deviation between them is accounted as the contribution of the counter-rotating-wave terms.
In the weak coupling regime, we observed that two QFIs about different parameters were monotonically decreased.
When the strength of the coupling become more stronger, the behavior of the non-Markovian could be observed from the QFIs perspective.

Finally, we generalized the results to the qudit system,
expressing the two QFIs in terms of the generalized Bloch vector.
Those expressions were valid for the $N$-qubit system with symmetry exchange.
We also considerd the QFI in the presence of the collisional dephasing with the initial state being prepared as a GHZ state.
The affine matrices for this dephasing process was derived and
the Bloch vector of the GHZ state also was expressed as the Bloch vector.
We found the dynamical QFI exponentially decrease by a factor of $e^{-2N^2\gamma t}$.

\section{Acknowledgments}

We would like to thank Dr. Xiao-Ming Lu for helpful advice.
XGW acknowledges support from the NFRPC
through Grant No. 2012CB921602 and the NSFC through
Grants No. 11025527 and No. 10935010.
FN acknowledges partial support from the LPS, NSA, ARO, NSF grant No. 0726909,
JSPS-RFBR contract No. 09-02-92114, Grant-in-Aid for Scientific Research (S),
MEXT Kakenhi on Quantum Cybernetics, and the JSPS through its FIRST program.
ZS acknowledges support from the National Nature Science Foundation of China with Grant No. 11005027;
the National Science Foundation of Zhejiang Province with Grant No. Y6090058,
and the Program for HNUEYT with Grant No. 2011-01-011.

\appendix

\section{Generators of the Lie algebra ${\rm \mathfrak{su}}\left(d\right)$}
\label{sec:AppI}
\makeatletter
\renewcommand \theequation{A.\@arabic \c@equation }
\makeatother \setcounter{equation}{0}

In group theory, the generators of the Lie algebra have the following properties:
i) Hermitian:
\begin{equation}
  \hat{\eta}_i^\dagger=\hat{\eta}_i;
\end{equation}
ii) traceless:
\begin{equation}
  {\rm Tr}\,\hat{\eta}_i=0;
\end{equation}
iii) orthogonality and normalization with respect to the trace metric relation:
\begin{equation}
  \frac{1}{2}\,{\rm Tr}\left(\hat{\eta}_{i}\,\hat{\eta}_{j}\right)=\delta_{ij}.
\end{equation}

Moreover, they also satisfy two relations as follows,
characterized by the structure constants $f_{ijk}$ and $g_{ijk}$,
\begin{eqnarray}
  \left[\hat{\eta}_i,\ \hat{\eta}_j\right] &=& 2\,i\sum_k\, f_{ijk}\,\hat{\eta}_k,\label{eq:Lie_pro1}\\
  \left\{\hat{\eta}_i,\ \hat{\eta}_j\right\} &=& \frac{4}{d}\,\delta_{ij}\,\openone_{d}+2\sum_k\, g_{ijk}\,\hat{\eta}_k,\label{eq:Lie_pro2}
\end{eqnarray}
where $\openone_{d}$ is the unit matrix of dimension $d$,
and $f_{ijk}$ ($g_{ijk}$) denotes the completely antisymmetric (symmetric) tensor.
The structure constants are determined by
\begin{eqnarray}
  f_{ijk} &=& \frac{1}{4\,i}\,{\rm Tr}\left(\left[\hat{\eta}_i,\hat{\eta}_j\right]\hat{\eta}_k\right),
  \label{eq:tensor_ant}\\
  g_{ijk} &=& \frac{1}{4}\,{\rm Tr}\left(\left\{\hat{\eta}_i,\hat{\eta}_j\right\}\hat{\eta}_k\right).
  \label{eq:tensor_sym}
\end{eqnarray}
Employing completeness relation of generators of ${\rm \mathfrak{su}}\left(d\right)$,
one can expand an arbitrary $d$-dimensional Hermitian matrix $X$ as
\begin{equation}
  X=\frac{1}{d}\,{\rm Tr}\left(X\right)\openone_{d}+\sum_{k=1}^{d^{2}-1}x_{k}\,\hat{\eta}_{k},\,x_{k}\in\mathbb{R},
\end{equation}
with $x_{k}={\rm Tr}\left(X\,\hat{\eta}_{k}\right)$.

Below, we systematically construct the generators $\mathbf{\hat{\eta}}=\left\{ \eta_{j}\right\} _{j=1}^{d^{2}-1}$
of the Lie algebra ${\rm \mathfrak{su}}\left(d\right)$ which are
given as follows \cite{Lendi1987,LendiBook2007}.
On a $d$-dimensional Hilbert space $\mathcal{H}\in\mathbb{C}^{d}$
spanned by an orthonormal set of states $\left\{\left|\,m\,\right\rangle\right\}_{m=1}^{d}$,
we first construct two sets of block off-diagonal Hermitian traceless matrices,
\begin{eqnarray}
  \hat{\mathcal{S}}_{m,n}
  &=& \left|m\right\rangle\left\langle n\right|+\left|n\right\rangle\left\langle m\right|, \label{eq:s_gen}\\
  \hat{\mathcal{A}}_{m,n}
  &=& -i\left(\left|m\right\rangle\left\langle n\right|-\left|n\right\rangle\left\langle m\right|\right),\label{eq:j_gen}
\end{eqnarray}
for $1\leq m<n\leq d$.
The above two sets contain the same number of elements as $\left(d^2-d\right)/2$.
Then, we construct $d-1$ real diagonal traceless matrices
\begin{equation}
  \hat{\mathcal{D}}_{k}=\sqrt{\frac{2}{k\left(k+1\right)}}
  \Big[\sum_{m=1}^{k}\left|m\right\rangle\left\langle m\right|-k\,\left|k+1\right\rangle\left\langle k+1\right|\Big],\label{eq:d_gen}\\
\end{equation}
for $\left(1\leq k\leq d-1\right)$.
Hitherto, the generators for the Lie algebra ${\rm \mathfrak{su}}\left(d\right)$
are obtained by the new set \cite{Lendi1987,LendiBook2007},
\begin{equation}
  \left\{\hat{\eta_{i}}\right\}_{i=1}^{d^{2}-1}=
  \left\{\hat{\mathcal{S}}_{m,n},\,\hat{\mathcal{A}}_{m,n},\,\hat{\mathcal{D}}_{k}\right\}.
  \label{eq:eta}
\end{equation}
The generators for $d=2$ coincide with the Pauli matrices with the structure constants
$f_{ijk}$ being Levi-Civita symbol $\epsilon_{ijk}$ and $g_{ijk}=0$.
The generators of ${\rm \mathfrak{su}}\left(3\right)$ are described by the Gell-Mann matrices given by \cite{LendiBook2007}
\begin{eqnarray}
 \hat{\eta}_{1} = \left(\begin{array}{ccc}
0 & 0 & 1\\
0 & 0 & 0\\
1 & 0 & 0
\end{array}\right);&
 \hat{\eta}_{2} = \left(\begin{array}{ccc}
0 & 1 & 0\\
1 & 0 & 0\\
0 & 0 & 0
\end{array}\right);\nonumber\\
 \hat{\eta}_{3} = \left(\begin{array}{ccc}
0 & 0 & 0\\
0 & 0 & 1\\
0 & 1 & 0
\end{array}\right); &
 \hat{\eta}_{4} = \left(\begin{array}{ccc}
0 & 0 & -i\\
0 & 0 & 0\\
i & 0 & 0
\end{array}\right);\nonumber\\
 \hat{\eta}_{5} = \left(\begin{array}{ccc}
0 & -i & 0\\
i & 0 & 0\\
0 & 0 & 0
\end{array}\right);&
 \hat{\eta}_{6} = \left(\begin{array}{ccc}
0 & 0 & 0\\
0 & 0 & -i\\
0 & i & 0
\end{array}\right);\nonumber\\
 \hat{\eta}_{7} = \left(\begin{array}{ccc}
1 & 0 & 0\\
0 & -1 & 0\\
0 & 0 & 0
\end{array}\right);&
 \hat{\eta}_{8} = \frac{1}{\sqrt{3}}\left(\begin{array}{ccc}
1 & 0 & 0\\
0 & 1 & 0\\
0 & 0 & -2
\end{array}\right).
\end{eqnarray}

\section{Derivation of equation (\ref{eq:QFI1_qudit})}
\label{sec:AppII}
\makeatletter
\renewcommand \theequation{B.\@arabic \c@equation }
\makeatother \setcounter{equation}{0}

Now, we provide the detailed derivation of the first line of Eq.~\eqref{eq:QFI1_qudit}.
In order to express $\mathcal{F}_{\lambda}$ in terms of the Bloch vector,
We should firstly write all the Hermitian operator in the same representation,
expanding them over the common generators of the Lie algebra $\mathfrak{su}(d)$ \cite{Ueda2010}.
The density matrix $\rho$ is described by Eq.~\eqref{eq:Bloch_rep},
and the SLD operator is supposed to be expanded as
\begin{equation}
  L=a\openone_{d}+\bm{b}\cdot\hat{\bm{\eta}},\label{eq:app_SLD}
\end{equation}
where $a$ and $\bm{b}$ are respectively real number and real vector to be determined.

Suppose $\rho$ contains some unknown parameter $\lambda$,
i.e., the Bloch vector $\bm{\omega}$ is $\lambda$-dependent.
Then we have
\begin{equation}
  \partial_{\lambda}\rho=\frac{1}{2}\left(\partial_{\lambda}\bm{\omega}\right)^{{\rm T}}\hat{\bm{\eta}}.\label{eq:app_partial_rho}
\end{equation}
By substituting Eqs.~\eqref{eq:app_SLD} and \eqref{eq:app_partial_rho} into Eq.~\eqref{eq:QFI1},
we find that $\mathcal{F}_{\lambda}$ can be rewritten as
\begin{equation}
  \mathcal{F}_{\lambda}
  =\left(\partial_{\lambda}\bm{\omega}\right)^{\rm T}\bm{b}
  \label{eq:QFI1_vecb}.
\end{equation}
Here, we use Eq.~\eqref{eq:qudit_Tr_relation} and the traceless property of generators.
Equation~\eqref{eq:QFI1_vecb} shows that when $\bm{b}$ is derived,
then the problem is solved.

To determine $\bm{b}$, we should use Eq.~\eqref{eq:SLD}.
Furthermore, by using Eq.~\eqref{eq:Lie_pro2},
we have following relation
\begin{eqnarray}
  \left\{\bm{a}\cdot\hat{\bm{\eta}},\,\bm{b}\cdot\hat{\bm{\eta}}\right\}
  &=& \sum_{ij}a_{i}b_{j}\left\{\hat{\eta}_i,\,\hat{\eta}_j\right\} \nonumber\\
  &=& \frac{4}{d}\,\bm{a}^{\rm T}\bm{b}\,\openone_{d}+2\sum_{ijk}\,g_{ijk}a_{i}b_{j}\,\hat{\eta}_k,
  \label{eq:app_anticomm}
\end{eqnarray}
With the help of Eqs.~\eqref{eq:Bloch_rep}, \eqref{eq:app_SLD}, and \eqref{eq:app_anticomm},
the right-hand side (RHS) of Eq.~\eqref{eq:SLD} reads
\begin{eqnarray*}
  \frac{1}{2}\left\{\rho,\,L\right\}
  &=& \frac{1}{2}\Big(\{\frac{\openone}{d},\,a\openone+\bm{b}\cdot\hat{\bm{\eta}}\}+ \nonumber\\
  & & \{\frac{1}{2}\bm{\omega}\cdot\hat{\bm{\eta}},\,a\openone\}
  +\{\frac{1}{2}\bm{\omega}\cdot\hat{\bm{\eta}},\,\bm{b}\cdot\hat{\bm{\eta}}\}\Big) \nonumber\\
  &=& \frac{1}{d}\left(a+\bm{\omega}^{{\rm T}}\bm{b}\right)\openone+ \nonumber\\
  & & \sum_{k}\Big(\frac{1}{d}b_{k}+\frac{1}{2}a\omega_{k}+\frac{1}{2}\sum_{ij}g_{ijk}\omega_{i}b_{j}\Big)\eta_{k}.
\end{eqnarray*}
The left-hand side (LHS) of Eq.~\eqref{eq:SLD} is given by Eq.~\eqref{eq:app_partial_rho}.
By comparing the terms on both side of Eq.~\eqref{eq:SLD},
one obtains
\begin{eqnarray}
  a+\bm{\omega}^{\rm T}\bm{b} &=& 0,\label{eq:openone_term} \\
  \left(\partial_{\lambda}\bm{\omega}\right)^{{\rm T}}\hat{\bm{\eta}}
  &=& \sum_{k}\Big(\frac{2}{d}b_{k}+a\omega_{k}+\sum_{ij}g_{ijk}\omega_{i}b_{j}\Big)\eta_{k} \nonumber\\
  &=& \frac{2}{d}\bm{b}^{{\rm T}}\hat{\bm{\eta}}
  +a\bm{\omega}^{{\rm T}}\hat{\bm{\eta}}+\sum_{ijk}g_{ijk}\omega_{i}b_{j}\hat{\eta}_{k} \nonumber\\
  &=& \frac{2}{d}\bm{b}^{{\rm T}}\hat{\bm{\eta}}
  +a\bm{\omega}^{{\rm T}}\hat{\bm{\eta}}+\sum_{jk}b_{j}G_{jk}\hat{\eta}_{k} \nonumber\\
  &=& \frac{2}{d}\bm{b}^{{\rm T}}\hat{\bm{\eta}}
  +a\bm{\omega}^{{\rm T}}\hat{\bm{\eta}}+\bm{b}^{{\rm T}}G\hat{\bm{\eta}}. \label{eq:eta_term}
\end{eqnarray}
Here, the matrix element of $G$ in Eq.~\eqref{eq:eta_term}
is given by $G_{jk}=\sum_{i}g_{ijk}\omega_{i}$ satisfying $G_{jk}=G_{kj}$.
With Eq.~\eqref{eq:openone_term}, we have
\begin{equation}
  a=-\bm{\omega}^{\rm T}\bm{b}=-\bm{b}^{\rm T}\bm{\omega}.
\end{equation}
By insetting the above equation into Eq.~\eqref{eq:eta_term},
we obtain
\begin{eqnarray}
\left(\partial_{\lambda}\bm{\omega}\right)^{{\rm T}}\hat{\bm{\eta}}
 &=& \bm{b}^{{\rm T}}\left(\frac{2}{d}-\bm{\omega}\bm{\omega}^{{\rm T}}+G\right)\hat{\bm{\eta}},\nonumber\\
 &=& \bm{b}^{{\rm T}}\mathcal{M}\hat{\bm{\eta}}, \label{eq:LinearEq}
\end{eqnarray}
by setting
\begin{equation*}
  \mathcal{M}\equiv\frac{2}{d}-\bm{\omega}\bm{\omega}^{{\rm T}}+G.
\end{equation*}
Hence, equation~\eqref{eq:LinearEq} directly gives the following equation
\begin{equation}
  \left(\partial_{\lambda}\bm{\omega}\right)^{{\rm T}}=\bm{b}^{{\rm T}}\mathcal{M}. \label{eq:LinerEq_sol}
\end{equation}
Suppose that we have the inverse matrix $\mathcal{M}^{-1}$.
After Eq.~\eqref{eq:F_omega}, we will make further discussions about $\mathcal{M}^{-1}$.
We have
\begin{equation}
  \bm{b}=\mathcal{M}^{-1}\left(\partial_{\lambda}\bm{\omega}\right) \label{eq:vecb}.
\end{equation}
from Eq.~\eqref{eq:LinerEq_sol}.
Finally, inserting Eq.~\eqref{eq:vecb} into Eq.~\eqref{eq:QFI1_vecb} yields
\begin{equation}
  \mathcal{F}_{\lambda}
  =\left(\partial_{\lambda}\bm{\omega}\right)^{{\rm T}}\mathcal{M}^{-1}\left(\partial_{\lambda}\bm{\omega}\right)\label{eq:F_omega}
\end{equation}

To calculate $\mathcal{F}_{\lambda}$,
we need to find the inverse of $\mathcal{M}$.
Generally, it may or may not exist,
i.e., $\mathcal{M}$ may have some zero eigenvalues.
In this case, we define $\mathcal{M}^{-1}$
on the support of $\mathcal{M}$, i.e., supp$\left(\mathcal{M}\right)$,
which is defined as a space spanned by those eigenvectors
with nonzero eigenvalues.
It is reasonable to do so.
By inserting Eq.~\eqref{eq:LinerEq_sol} into Eq.~\eqref{eq:QFI1_vecb},
we rewrite $\mathcal{F}_{\lambda}$ as
\begin{equation}
  \mathcal{F}_{\lambda}=\bm{b}^{{\rm T}}\mathcal{M}\bm{b}.\label{eq:QFI1_bmb}
\end{equation}
Suppose $\mathcal{M}$ has spectral decomposition
\begin{equation}
  \mathcal{M}=\sum_{i=1}^{d^2-1}m_{i}\bm{v}_{i}\bm{v}_{i}^{\rm T},\label{eq:M_spec}
\end{equation}
where $\bm{v}_{i}$ denotes eigenvector with eigenvalue $m_{i}$.
We assume $m_{i}\neq0$ for $i=1,...,n$ and
$m_{i}=0$ for $i=n+1,...,d^2-1$.
With Eq.~\eqref{eq:M_spec}, equation~\eqref{eq:QFI1_bmb} reads
\begin{equation}
  \mathcal{F}_{\lambda}=\sum_{i=1}^{n}m_{i}\bm{b}^{\rm T}\bm{v}_{i}\bm{v}_{i}^{\rm T}\bm{b}
\end{equation}
It is shown that $\mathcal{F}_{\lambda}$ is defined on
supp$\left(\mathcal{M}\right)$.

\section{Derivation of equation (\ref{eq:QFI2_qudit})}
\label{sec:AppIII}
\makeatletter
\renewcommand \theequation{C.\@arabic \c@equation }
\makeatother \setcounter{equation}{0}

In this appendix, we give the detailed derivation of $\mathcal{I}_{\lambda}$
in terms of the Bloch vector for mixed states given by Eq.~\eqref{eq:QFI2_qudit}.
We first expand the Hermitian matrix as
\begin{equation}
  \sqrt{\rho}=y\,\openone_{d}+\bm{x}\cdot\hat{\bm{\eta}},\label{eq:sqrt_rho}
\end{equation}
where $y={\rm Tr}\left(\sqrt{\rho}\right)/d$ and $\bm{x}$ is an unknown $d^2-1$ dimensional real vector.
With Eq.~\eqref{eq:sqrt_rho}, $\mathcal{I}_{\lambda}$ in Eq.~\eqref{eq:QFI2} can be rewritten as
\begin{equation}
  \mathcal{I}_{\lambda}=8\left[\left(\partial_{\lambda}y\right)^2+\frac{2}{d}\left|\partial_{\lambda}\bm{x}\right|^2\right],
  \label{eq:QFI2_qudit_mix}
\end{equation}
by using the properties of the generators of the Lie algebra given in App.~\ref{sec:AppI}.
Furthermore, we have the following equation
\begin{equation}
  \rho=\left(\sqrt{\rho}\right)^2.\label{eq:rho_sqrt2}
\end{equation}
In the Bloch representation, the left-hand side of the equation above is given by Eq.~\eqref{eq:Bloch_rep}, and
the right-hand side reads
\begin{equation}
  \left(\sqrt{\rho}\right)^2
  =\left(y^{2}+\frac{2}{d}\left|\bm{x}\right|^{2}\right)\openone_{d^{2}-1}+\sum_{k}\left(2yx_{k}+g_{ijk}x_{i}x_{j}\right)\hat{\eta}_{k}.\nonumber
\end{equation}
By comparing the terms on both sides, one finds that $y$ and $\bm{x}$ are completely determined
by the following $d^2$ quadratic nonlinear equations:
\begin{eqnarray}
  y^2+\frac{2}{d}\left|\bm{x}\right|^2 &=& \frac{1}{d},\\
  \sum_{i,j=1}^{d^2-1}g_{ijk}\,x_{i}\,x_{j}+2\,y\,x_{k} &=& \frac{\omega_{k}}{2},
\end{eqnarray}
for $k=1,2,...,d^2-1$.

Below, we give the detailed derivation for Eqs.~\eqref{eq:y_sol} and \eqref{eq:x_sol}.
Multiplying the LHS of Eq.~\eqref{eq:xy_fun2} by $4\,y\,\bm{x}^{{\rm T}}$ and
the RHS by $\bm{\omega}^{{\rm T}}$, we obtain
\begin{equation}
  16\,y^{2}\left|\bm{x}\right|^{2}=\left|\bm{\omega}\right|^{2}.
\end{equation}
By replacing $\left|\bm{x}\right|^{2}$ in equation above by
\begin{equation}
  \left|\bm{x}\right|^{2}=\frac{1}{2}-y^{2},
\end{equation}
given by Eq.~\eqref{eq:xy_fun1}, we get the following equation
\begin{equation}
  16\,y^{4}-8y^{2}+\left|\bm{\omega}\right|^{2}=0.
\end{equation}
Solving the above equation, we obtain two solutions
\begin{eqnarray}
  y_{+}^{2} & = & \frac{1+\sqrt{1-\left|\bm{\omega}\right|^{2}}}{4}, \label{eq:yp} \\
  y_{-}^{2} & = & \frac{1-\sqrt{1-\left|\bm{\omega}\right|^{2}}}{4}. \label{eq:ym}
\end{eqnarray}

We suppose that $\rho$ has spectral decomposition
\begin{equation}
  \rho=\sum_{i}\varrho_{i}\left|\psi_{i}\right\rangle \left\langle \psi_{i}\right|=\varrho_{1}\left|\psi_{1}\right\rangle \left\langle \psi_{1}\right|+\left(1-\varrho_{1}\right)\left|\psi_{2}\right\rangle \left\langle \psi_{2}\right|.
\end{equation}
Then we can derive the following inequality
\begin{equation}
  y=\frac{1}{2}{\rm Tr}\left(\sqrt{\rho}\right)=\frac{1}{2}\sum_{i}\sqrt{\varrho_{i}}\geq\frac{1}{2}.
\end{equation}
Then the solution $y_{-}^{2}$ in Eq.~\eqref{eq:ym} can be neglected.
Due to $y$ being nonnegative real number, we finally get Eq.~\eqref{eq:y_sol} from
Eq.~\eqref{eq:yp}.
Furthermore, substituting Eq.~\eqref{eq:y_sol} into Eq.~\eqref{eq:xy_fun2} directly
yields Eq.~\eqref{eq:x_sol}.

\section{Derivation of the affine transformation matrix $A$ and $\bm{c}$ for the collisional dephasing}
\label{sec:AppIV}
\makeatletter
\renewcommand \theequation{D.\@arabic \c@equation }
\makeatother \setcounter{equation}{0}

We firstly verify the elements of the Bloch vector
can be directly written out from the elements of the
density matrix.
A $d$-dimensional density matrix
is described by the generalized Bloch vector in
Eq.~\eqref{eq:Bloch_rep} in the Bloch representation.
The elements of the generalized Bloch vector is defined
by
\begin{equation}
  \omega_{i}={\rm Tr}\left(\rho\,\hat{\eta}_{i}\right). \label{eq:omega_i}
\end{equation}
In the appendix A, we systematically construct the
generators of Lie algebra ${\rm \mathfrak{su}}\left(d\right)$
in Eq.~\eqref{eq:eta}, which is defined by three sets
given by Eqs.~\eqref{eq:s_gen}, \eqref{eq:j_gen}, and \eqref{eq:d_gen}.
Then $\bm{\omega}$ also can be divided into three parts as
\begin{equation}
  \left\{\omega_{i}\right\}_{i=1}^{d^2-1}=
  \left\{\omega^{\mathcal{S}}_{m,n},\,\omega^{\mathcal{A}}_{m,n},\,\omega^{\mathcal{D}}_{k}\right\}.
\end{equation}
We find that the elements of the Bloch vector are directly given by the elements of the density matrix
\begin{eqnarray}
  \omega^{\mathcal{S}}_{m,n} &=& {\rm Tr}\left(\rho\,\hat{\mathcal{S}}_{m,n}\right)=2\,\Re\left(\rho_{m,n}\right),\label{eq:omega_s}\\
  \omega^{\mathcal{A}}_{m,n} &=& {\rm Tr}\left(\rho\,\hat{\mathcal{A}}_{m,n}\right)=-2\,\Im\left(\rho_{m,n}\right),\label{eq:omega_a}\\
  \omega^{\mathcal{D}}_{k} &=& {\rm Tr}\left(\rho\,\hat{\mathcal{D}}_{k}\right) \nonumber\\
  &=& \sqrt{\frac{2}{k\left(k+1\right)}}\Big[\sum_{m=1}^{k}\rho_{m,m}-k\,\rho_{k+1,k+1}\Big].\label{eq:omega_d}
\end{eqnarray}

The master equation of the collisional dephasing model is
described by Eq.~\eqref{eq:master_eq},
and the time evolution of the density matrix elements is
given by Eq.~\eqref{eq:master_eq result}.
With Eqs.~\eqref{eq:master_eq result}, \eqref{eq:omega_s},
\eqref{eq:omega_a}, and \eqref{eq:omega_s},
the elements of the affine-mapped Bloch vector is given by
\begin{eqnarray}
  \omega^{\mathcal{S}}_{m,n}\left(t\right)
  &=& e^{-\left(m-n\right)^{2}\gamma t}\,2\,\Re\left[\rho_{m,n}\left(0\right)\right]\nonumber\\
  &=& e^{-\left(m-n\right)^{2}\gamma t}\omega^{\mathcal{S}}_{m,n}\left(0\right),\label{eq:omega_s_t}\\
  \omega^{\mathcal{A}}_{m,n}\left(t\right)
  &=& -e^{-\left(m-n\right)^{2}\gamma t}\,2\,\Im\left[\rho_{m,n}\left(0\right)\right]\nonumber\\
  &=& e^{-\left(m-n\right)^{2}\gamma t}\omega^{\mathcal{A}}_{m,n}\left(0\right),\label{eq:omega_a_t}\\
  \omega^{\mathcal{D}}_{k}\left(t\right) &=& \omega^{\mathcal{D}}_{k}\left(0\right).\label{eq:omega_d_t}
\end{eqnarray}
Equation~\eqref{eq:omega_d_t} illustrates that
$\bm{\omega}^{\mathcal{D}}_{k}$ remains unchanged
under the collsional dephasing, since $\bm{\omega}^{\mathcal{D}}_{k}$
depend on the diagonal elements of the density matrix
which left unchanged by dephasing.
Writing in the form of affine map in Eq.~\eqref{eq:affine_map_qudit},
we finally obtain $A$ in Eq.~\eqref{eq:A_matrix} and $\bm{c}=0$
from Eqs.~\eqref{eq:omega_s_t}, \eqref{eq:omega_a_t}, and \eqref{eq:omega_d_t}.


\end{document}